\documentclass{article}
\usepackage[utf8]{inputenc}
\usepackage{hyperref}
\usepackage{amsmath}
\usepackage{braket}
\usepackage{footmisc}
\usepackage{graphicx}
\usepackage{dsfont}

  \usepackage[
    backend=biber,
    style=authoryear,
  ]{biblatex}
\usepackage[affil-it]{authblk} 
\usepackage{etoolbox}
\usepackage{lmodern}
\makeatletter
\patchcmd{\@maketitle}{\LARGE \@title}{\fontsize{20}{19.2}\selectfont\@title}{}{}
\makeatother

\title{Consciousness and the Collapse of the Wave Function\thanks{Forthcoming in (S. Gao, ed.) \textit{Consciousness and Quantum Mechanics} (Oxford University Press).  Authors are listed in alphabetical order and contributed equally.  We owe thanks to audiences starting in 2013 at Amsterdam, ANU, Cambridge, Chapman, CUNY, Geneva, Göttingen, Helsinki, Mississippi, Monash, NYU, Oslo, Oxford, Rio, Tucson, and Utrecht. These earlier presentations have occasionally been cited, so we have made some of them available at \href{http://consc.net/qm}{consc.net/qm}. For feedback on earlier versions, thanks to Jim Holt, Adrian Kent, Kobi Kremnizer, Oystein Linnebo,  and Trevor Teitel.  We are grateful to Maaneli Derakhshani and Philip Pearle for their help with the mathematics of collapse models, and especially to Johannes Kleiner, who coauthored section 5 on quantum integrated information theory.}} 
 
 \author[$\dagger$]{David J. Chalmers}
 \author[$\ddagger$]{Kelvin J. McQueen}
 \affil[$\dagger$]{New York University}
 \affil[$\ddagger$]{Chapman University}

\begin{document}

\maketitle

\begin{abstract}
    Does consciousness collapse the quantum wave function?  This idea was taken seriously by John von Neumann and Eugene Wigner but is now widely dismissed.  We develop the idea by combining a mathematical theory of consciousness (integrated information theory) with an account of quantum collapse dynamics (continuous spontaneous localization).  Simple versions of the theory are falsified by the quantum Zeno effect, but more complex versions remain compatible with empirical evidence.   In principle, versions of the theory can be tested by experiments with quantum computers.  The upshot is not that consciousness-collapse interpretations are clearly correct, but that there is a research program here worth exploring.
    \end{abstract}


\tableofcontents

\newpage

\section{Introduction}

One of the hardest philosophical problems arising from contemporary science is the problem of quantum reality. What is going on in the physical reality underlying the predictions of quantum mechanics?  It is widely accepted that quantum-mechanical systems are describable by a wave function. The wave function need not assign definite position, momentum, and other definite properties to physical entities.   Instead it may assign a superposition of multiple values for position, momentum, and other properties. When one measures these properties, however, one always obtains a definite result.  On a common picture, the wave function is guided by two separate principles.   First, there is a process of evolution according to the Schrödinger equation, which is linear, deterministic, and constantly ongoing.  Second, there is a process of collapse into a definite state, which is nonlinear, nondeterministic, and happens only on certain occasions of measurement.

This picture is standardly accepted at least as a basis for empirical predictions, but it has been less popular as a story about the underlying physical reality. The biggest problem is the \textit{measurement problem} (see \textcite{Albert1992}; \textcite{Bell1990}). On this picture, a fundamental measurement-collapse principle says that collapses happen when and only when a measurement occurs.  But on the face of it, the notion of “measurement” is vague and anthropocentric, and is inappropriate to play a role in a fundamental specification of reality. To make sense of quantum reality, one needs a much clearer specification of the underlying dynamic processes.

Another of the hardest philosophical problems arising from contemporary science is the mind-body problem. What is the relation between mind and body, or more specifically, between consciousness and physical processes?  By consciousness, what is meant is phenomenal consciousness, or subjective experience.  A system is conscious when there is something it is like to be that system, from the inside.  A mental state is conscious when there is something it is like to be in that state.

There are many aspects to the problem of consciousness, including the core problem of why physical processes should give rise to consciousness at all.  One central aspect of the problem is the \textit{consciousness-causation problem}: how does consciousness play a causal role in the physical world? It seems obvious that consciousness plays a causal role, but it is surprisingly hard to make sense of what this role is and how it can be played.

There is a long tradition of trying to solve the consciousness-causation problem and the quantum measurement problem at the same time, by saying that measurement is an act of consciousness, and that consciousness plays the role of bringing about wave function collapse. The locus classicus of this consciousness-collapse thesis is Eugene Wigner's 1961 article “Remarks on the mind-body question”. There are traces of the view in earlier work by \textcite{vonNeumann1955} and \textcite{LondonBauer1939}.\footnote{It is clear that \textcite{vonNeumann1955} endorses a measurement-collapse interpretation, and he says (p.418) that subjective perception is “related” to measurement, but he does not clearly identify measurement with conscious perception.  In his discussion of observed systems (I), measuring instruments (II), and “actual observer” (III), he says “the boundary can just as well be drawn between I and II+III as between I+III and III”.  This suggests neutrality on whether the collapse process is triggered by measuring devices or by conscious observers.   He also says that the boundary is “arbitrary to a very large extent” (p.420), which is not easy to reconcile with the fact that different locations for collapse are empirically distinguishable in principle, as we discuss in section 6. \textcite[section 11]{LondonBauer1939} say more clearly: “We note the essential role played by the consciousness of the observer in this transition from the mixture to the pure case.  Without his effective intervention, one would never obtain a new psi function” (although see \textcite{French2020} for an alternative reading).} In recent years the approach has been pursued by Henry Stapp (1993) and others.

The central motivations for the consciousness-collapse view come from the way it addresses these problems.  Where the problem of quantum reality is concerned, the view provides one of the few interpretations of quantum mechanics that takes the standard measurement-collapse principle at face value.  Other criteria for measurement may be possible, but understanding measurement in terms of consciousness has a number of motivations.  First, it provides one of the few non-arbitrary criteria for when measurement occurs.  Second, it is arguable that our core pretheoretical concept of measurement is that of measurement by a conscious observer.  Third, the consciousness-collapse view is especially well-suited to save the central epistemological datum that ordinary conscious observations have definite results. Fourth, understanding measurement as consciousness provides a potential solution to the consciousness-causation problem: consciousness causes collapse. 

Despite these motivations, the consciousness-collapse view has not been popular among contemporary researchers in the foundations of physics. Some of this unpopularity may stem from the popularity of the view in unscientific circles: for example, popular treatments by \textcite{Capra1975} and \textcite{Zukav1979}, who link the view to Eastern religious traditions. More substantively, the view is frequently set aside in the literature on the basis of \textit{imprecision} and on the basis of \textit{dualism}.

The objection from imprecision is stated succinctly by   \textcite[pp.82--3]{Albert1992}

\begin{quote}
    “How the physical state of a certain system evolves (on this proposal) depends on whether or not that system is conscious; and so in order to know precisely how things physically behave, we need to know precisely what is conscious and what isn't. What this “theory” predicts will hinge on the precise meaning of the word conscious; and that word simply doesn't have any absolutely precise meaning in ordinary language; and Wigner didn't make any attempt to make up a meaning for it; so all this doesn't end up amounting to a genuine physical theory either.”
\end{quote}

We think that the force of this objection is limited. Of course it is true that ‘conscious’ in ordinary language is highly ambiguous and imprecise, but it is easy to disambiguate the term and make it more precise.  Philosophers have distinguished a number of meanings for the term, the most important of which is phenomenal consciousness. As usually understood, a system is phenomenally conscious when there is something it is like to be that system: so if there is something it is like to be a bat, a bat is phenomenally conscious, and if there is nothing it is like to be a rock, a rock is not phenomenally conscious.  One might question the precision of this concept in turn, but it is at least a common and widely defended view (see e.g. \textcite{Antony2006}; \textcite{Simon2017}) that it picks out a definite and precise property.  On this view, phenomenal consciousness comes in a number of varieties, but it is either definitely present or definitely absent in a given system at a given time.

In recent years, theories that give precise mathematically-defined conditions for the presence or absence of consciousness have begun to be developed.  The most well-known of these theories is Tononi’s integrated information theory (\cite{Tononi2008}), which specifies a mathematical structure for conscious states and quantifies them with a mathematical measure of integrated information.  Of course it is early days in the science of consciousness, and current theories are unlikely to be final theories.  Nevertheless, it is possible to envisage precise theories of consciousness, and to reason about they might be combined with a consciousness-collapse view to yield precise interpretations of quantum mechanics.  

Crucially, when different precise theories of consciousness are combined with the consciousness-collapse view, these yield subtly different experimental predictions. As a result, we have a further motivation for taking consciousness-collapse interpretations seriously: they can be tested experimentally.  As we discuss in section \ref{experiments}, there is a long-term research program of experimentally testing consciousness-collapse interpretations and eventually supporting a precise consciousness-collapse interpretation.  The required experiments are difficult, but advances in quantum computing may already exclude certain simple consciousness-collapse interpretations. Because of these considerations, the underdetermination of conditions for consciousness does not reflect any fundamental imprecision in consciousness-collapse views.  It simply reflects an experimentally testable degree of freedom.  

The second common objection to the consciousness-collapse view is that it is committed to dualism: the view that the mental and the physical are fundamentally distinct. The consciousness-collapse view treats consciousness in a special way that seems to exempt it from the standard quantum-mechanical laws governing physical systems.  This remark by Peter Lewis (this volume) reflects a common attitude:

\begin{quote}
    “Wigner postulates a strong form of interactive dualism in order to justify a duality in the physical laws. Few will want to follow Wigner down this path: non-physical minds, especially causally active ones, are mysterious at best.”
\end{quote}

Again, we think the force of this objection is limited.

First: the consciousness-collapse thesis need not lead to dualism.  It is compatible with materialist views on which consciousness is a complex physical property.  For example, let us suppose a materialist version of integrated information theory on which consciousness is identical to $\Phi^{*}$, the property of having integrated information above a certain threshold.  Then the consciousness-collapse theory will say that $\Phi^{*}$ causes collapse. This interpretation of quantum mechanics will involve a fundamental physical law saying that under the conditions specified by $\Phi^{*}$, collapse is brought about according to the Born rule.  A fundamental law involving a complex physical property may be unlike familiar physical laws, but it involves nothing nonphysical.

Second: where consciousness is concerned, there are reasons to take dualism seriously. There are familiar reasons to question whether any purely physical theory can explain consciousness.  One common reason (\cite{Chalmers2003}) is that physical theories explain only structure and dynamics (the so-called “easy problems” of behavior and the like), and explaining consciousness (the so-called “hard problem”) requires explaining more than structure and dynamics. These reasons need not lead to substance dualism, on which consciousness involves a separate nonphysical entity akin to an ego or soul, but they have led many theorists to adopt a form of property dualism where consciousness is accepted as a fundamental property akin to spacetime, mass, and charge.

Where physical theories give fundamental physical laws that connect physical properties to each other, a property dualist theory of consciousness gives fundamental psychophysical laws that connect physical properties to consciousness. For example, on a property dualist construal of integrated information theory, there might be a fundamental physics-to-consciousness law saying that when a system has $\Phi$ above a certain threshold, the system will have a corresponding state of consciousness. Such a law has a structure akin to the Newtonian mass-to-gravitational-field law, saying that when a system has a certain mass, the system will have a corresponding gravitational field. On a consciousness-causes-collapse theory, there will be an additional consciousness-to-physics law saying that states of consciousness bring about wave function collapse in a certain way. Putting these theories together might yield a mathematically precise version of property dualism that specifies the conditions under which consciousness arises and the role that it plays.

Interestingly, the most common reason among philosophers for rejecting property dualist theories of consciousness is an argument from physics. This argument runs roughly as follows: (1) every physical effect has only physical causes, (2) consciousness causes physical effects, so (3) consciousness is physical.  The key first premise is a causal closure thesis, supported by the observation that there are no causal gaps in standard physics that a nonphysical consciousness might fill.  But wave function collapse in quantum mechanics appears to be precisely such a gap, and consciousness-collapse models are at least not obviously ruled out by known physics.  The situation is that many physicists rule out consciousness-collapse models for philosophical reasons (they are dualistic), while philosophers rule out property dualist models for physics-based reasons (they violate causal closure).

The upshot is that a central reason to reject the consciousness-collapse thesis (it leads to dualism) and a central reason to reject interactionist property dualism (it violates the causal closure of physics) provide no reason to reject the two views when taken together.  Perhaps there are other reasons to reject the consciousness-collapse thesis or to reject dualism, but these reasons must be found elsewhere.

A third common objection to the consciousness-collapse thesis is that it is not \textit{necessary} to invoke consciousness in an interpretation of quantum mechanics, as there are alternative interpretations that give it no special role.  Even if we retain the measurement-collapse framework, it is possible to understand measurement independently of consciousness, so that nonconscious systems such as ordinary measuring devices can collapse the wave function.  Going beyond this framework, a number of alternative interpretations have been developed that give no role to the notion of measurement.  These include spontaneous-collapse interpretations (e.g. \textcite{Pearle1976}; \textcite{Ghirardi1986}) which retain a collapse process but dispense with the need for measurement as a trigger, and hidden-variable interpretations (\cite{Bohm1952}) and many-worlds interpretations (\cite{Everett1957}), which eliminate collapse entirely.

We agree that one is not forced to accept a role for consciousness in quantum mechanics.  At the same time, the mere existence of alternative interpretations is not itself good reason to reject the consciousness-collapse thesis.  If it were, we would have good reason to reject all interpretations.  Perhaps the underlying thought is that the consciousness-collapse thesis is extravagant and has certain costs, such as dualism. For there to be a serious objection here, an opponent needs to articulate the costs as objections in their own right.  As with every other interpretation of quantum mechanics, the consciousness-collapse interpretation has both serious costs (dualism) and serious benefits (taking the standard dynamics at face value, solving the consciousness-causation problem). To assess any interpretation, we need to weigh its costs against its benefits.

In this article, we are exploring consciousness-collapse models rather than endorsing them. In particular, we are not asserting that these interpretations are superior to other interpretations of quantum mechanics.  Both of us have considerable sympathy with other interpretations and especially with many-worlds interpretations (see \textcite[ch.10]{Chalmers1996} and \textcite{McQueenVaidman2019}).  But we think that consciousness-collapse interpretations deserve close attention.  If it turns out that these interpretations have fatal flaws, they can be set aside.  But if there are consciousness-collapse interpretations without clear fatal flaws, then these interpretations should be taken seriously as possible descriptions of quantum-mechanical reality.

In our view, by far the most important challenge to consciousness-collapse models is not the issue of imprecision or of dualism, but the question of \textit{dynamic principles}.  Can we find a simple, coherent, and empirically viable set of dynamic principles governing how consciousness collapses the wave function?  If we can find such principles, consciousness-collapse models should be placed alongside other dynamic models (including Bohmian hidden-variable models, Everettian many-worlds models, and Pearle-GRW style spontaneous collapse models) as serious contenders to be the correct interpretation of quantum mechanics. If we cannot, then consciousness-collapse models may remain an important speculative class of models, but they will stay on the second tier of interpretations until they are cashed out with dynamic principles.

In what follows, we will explore the prospects for consciousness-collapse interpretations of quantum mechanics.  We will do this mainly by exploring and evaluating potential dynamic principles. We focus especially on what we call \textit{super-resistance models}, according to which there are special properties that resist superposition and trigger collapse.  When these models are combined with the consciousness-collapse thesis, we obtain models in which consciousness or its physical correlates resist superposition and trigger collapse. We think super-resistance consciousness-collapse models are worth investigating, and in this article we investigate some of them. 

In this article we are not trying to solve the hard problem of how physical processes give rise to consciousness.  We are giving an account of the causal role of consciousness that can be combined with many different approaches to the hard problem.  Our approach is consistent with both materialist views, on which consciousness is identified with a complex physical property, and dualist views, on which consciousness is a primitive property that correlates with physical properties.  Our approach is also consistent with many different theories of consciousness that correlate consciousness with underlying physical processes.  For concreteness we will often assume a Tononi-style theory of consciousness on which consciousness is identical to or correlated with integrated information, but much of what we say should translate straightforwardly to other theories of consciousness.

We will not be addressing problems that come up for collapse models of quantum mechanics quite generally.  For example, collapse models face important challenges stemming from the theory of relativity (collapse seems to require a privileged reference frame (\cite{Maudlin2011})), and the tails problem (collapse leaves wave functions with tails (\cite{McQueen2015})).  The collapse models we consider certainly face these challenges.  These are important challenges, but for present purposes we will be happy if consciousness-collapse interpretations can be shown to be about as viable as widely discussed spontaneous-collapse interpretations.  Interpretations in both classes will still face the general problems. A number of ideas about how to deal with them have been put forward, but this is a topic for another day.

Our aim is to set out the best consciousness-collapse model that we can and to assess it.  Our discussion is speculative and our conclusions are mixed.  We articulate both positive models and serious limitations.  We first articulate a simple consciousness-collapse model on which consciousness is entirely superposition-resistant.  This model is subject to a conclusive objection (distinct from those outlined above) arising from the quantum Zeno effect.  We then articulate a model that is not subject to this objection, combining integrated information theory with Pearle’s continuous-collapse theory. We explore the prospects of empirically testing these models, and discuss some objections.  The model is still subject to both empirical and philosophical objections, but there are some potential ways forward.  The upshot is not that consciousness-collapse interpretations are clearly correct, but that there is a research program here worth exploring.

\section{Consciousness as super-resistant}

One can clarify the options for a consciousness-collapse theory by asking a crucial question for any collapse model of quantum mechanics: What is the locus of collapse?  That is, which observable determines the definite states that the collapse process projects superposed states onto?  Here there are two options: there can be a \textit{variable locus} (different observables serve as the locus on different occasions of collapse) or a \textit{fixed locus} (the same observable always serves as the locus of collapse).

A variable-locus model is closest to standard formulations of quantum mechanics. On a standard understanding, many different observable quantities (e.g. position, momentum, mass, and spin) can be measured and thereby serve as the locus of collapse.  Every observable is associated with an operator. Upon measurement, the wave function collapses probabilistically into an eigenstate of that operator, and the measurement reveals the corresponding eigenvalue for the observable (such as a specific position for the particle), with probabilities determined by the prior quantum state according to the Born rule.

Henry Stapp’s consciousness-collapse model (\cite{Stapp1993}) is a variable-locus model, on which consciousness collapses whatever observable is being consciously observed at a given time.  The variable-locus approach has some attractions, but it also faces some hard questions.  Not least is the question: what determines which observable is being measured?  This question is hard enough that Stapp's model postulates an entirely separate process that determines the locus of collapse. Stapp calls this process “asking a question of nature”, which is supposed to be something that takes place in the mind of an observer. Stapp takes this to be a third process distinct from von Neumann’s standard dual processes of collapse itself and Schrödinger evolution.  Stapp takes this third process as primitive.  There are options for analyzing it (perhaps via a precisely specified observation relation between observers and observables, for example, or by building awareness of observables into the structure of consciousness), but it is clear that such a theory will be complex.

One option for a variable-locus consciousness-collapse theory invokes the idea that consciousness represents certain objects and properties in its environment.  For example, visual experiences typically represent the color, shape, and location of observed objects, while auditory experiences represent locations, pitches, and the like.  A consciousness-collapse view may hold that when consciousness represents observable properties of an observed object, the object collapses into a definite state of those observables.  For example, perceiving the location of a ball that was previously in a superposition will collapse the ball into a definite location. One trouble here is that on standard representationalist views, the represented properties are built into a state of consciousness but the represented objects are not.  In some cases an experience as of a single object may be caused by no object or by multiple objects in reality, so there is still a difficult question about which object if any undergoes collapse.  This approach may work better with relationist views where consciousness involves direct awareness of specific objects and properties, but there will still be many complications.\footnote{For representationalist views, see \textcite{Tye1995}.  For relationist views, see \textcite{ByrneLogue2009}.  These views may face a version of the Zeno problem in the next section, arising from whether the states of consciousness themselves can enter superpositions.}

Fixed-locus models are simpler in a number of respects, and we will focus on them.  In a fixed-locus measurement-collapse model, there are special properties that serve as the locus of collapse.  In a fixed locus consciousness-collapse model, consciousness itself (or perhaps its physical correlate) serves as the locus of collapse.  It is this idea that we will develop in what follows.

One natural way to develop a fixed-locus collapse model is through the idea of superposition-resistance, which we will sometimes abbreviate as super-resistance.  The idea is that there are special superposition-resistant observables, which as a matter of fundamental law resist superposition and cause the system to collapse onto eigenstates of these observables (with probabilities given by the Born rule). The corresponding class of models are \textit{super-resistance models} of quantum mechanics.\footnote{In earlier versions of this article we called superposition-resistant observables “m-properties” (short for “measurement properties”) and super-resistance models “m-property models”.} There are a number of different ways to make the dynamics of super-resistance precise, some of which we will explore in the following sections.  A strong version of super-resistance invokes fundamental \textit{superselection rules} (\cite{WickWightmanWigner1952}), according to which certain observables are entirely forbidden from entering superpositions.  A weaker version invokes principles according to which these superpositions are unstable and tend to collapse. 

There are super-resistance models of collapse that give no special role to consciousness or measurement.  One well-known super-resistance model is Penrose’s model (\cite{Penrose2014}) of quantum mechanics on which spacetime structure is superposition-resistant: when the structure of spacetime evolves into superpositions over a certain threshold, these superpositions collapse onto a definite structure.  One can also see the GRW interpretation of quantum mechanics as an interpretation on which position is mildly superposition-resistant: superpositions of position tend to collapse, though with low probability for isolated particles.

Super-resistance models work well with measurement-collapse interpretations of quantum mechanics.  In the context of these interpretations, we can think of a super-resistant property not as a \textit{measured} property (e.g. particle position) but as a \textit{measurement} property (e.g. a pointer position or a conscious experience). To sketch the idea intuitively: suppose there is a special class of measurement devices (e.g. oscilloscopes) which have special measurement properties (e.g. meter readings or pointer locations) that (as a matter of fundamental law) resist superposition and tend to collapse.  When a measurement takes place, a measured property affects a measurement property.  Suppose that we have a quantum system (e.g. a particle) in a superposition of locations $a$ and $b$, which we represent (simplifying by omitting amplitudes) as the quantum state $\ket{a} + \ket{b}$.  The particle interacts with a measurement system such that if not for this principle, it would yield an entangled superposition
 $\ket{a}\ket{M(a)} + \ket{b}\ket{M(b)}$, where $M(a)$ and $M(b)$ are the states of the measurement system. Because $M$ is superposition-resistant, the particle and measurement system will instead evolve into a collapsed state $\ket{a}\ket{M(a)}$ or $\ket{b}\ket{M(b)}$, with probabilities given by the Born rule.  The effect will be much the same as if the measured property collapsed directly, but now the measurement properties serve as a single locus of collapse.

Superposition-resistance is an especially natural idea in the context of consciousness-collapse models of quantum mechanics. The idea that consciousness resists superposition is suggested in a brief passage in \textcite{Wigner1961}, and is later developed by \textcite{Albert1992}, and \textcite{Chalmers2003}.

Wigner writes:

\begin{quote}
    ``If the atom is replaced by a conscious being, the wave function $\alpha(\phi_1 \times \chi_1) + \beta (\phi_2 \times \chi_2)$ (which also follows from the linearity of the equations) appears absurd because it implies that my friend was in a state of suspended animation before he answered my question. It follows that the being with a consciousness must have a different role in quantum mechanics than the inanimate measuring device: the atom considered above. In particular, the quantum mechanical equations of motion cannot be linear.” (\cite[p.180]{Wigner1961})
    \end{quote}
    
Wigner’s suggestion seems to be that a state of consciousness cannot be superposed because it would require being in a “state of suspended animation”.  Wigner does not suggest a dynamic process for collapse here, but potential processes are fleshed out a little by Albert and Chalmers.  Albert suggests a picture on which the physical correlates of consciousness immediately collapse once superposed:

    \begin{quote}
        All physical objects almost always evolve in strict accordance with the dynamical equations of motion. But every now and then, in the course of some such dynamical evolutions (in the course of measurements, for example), the brain of a sentient being may enter a state wherein (as we've seen) states connected with various different conscious experiences are superposed; and at such moments, the mind connected with that brain (as it were) opens its inner eye, and gazes on that brain, and that causes the entire system (brain, measuring instrument, measured system, everything) to collapse, with the usual quantum-mechanical probabilities, onto one or another of those states; and then the eye closes, and everything proceeds again in accordance with the dynamical equations of motion until the next such superposition arises, and then that mind's eye opens up again, and so on. (\cite[pp.81-2]{Albert1992})
    \end{quote}

Albert is entertaining the view mainly for the sake of argument, and he almost immediately rejects it in the passage quoted earlier about the imprecision of consciousness.  Chalmers writes more sympathetically:

    \begin{quote}
        Upon observation of a superposed system, Schrödinger evolution at the moment of observation would cause the observed system to become correlated with the brain, yielding a resulting superposition of brain states and so (by psychophysical correlation) a superposition of conscious states. But such a superposition cannot occur, so one of the potential resulting conscious states is somehow selected (presumably by a nondeterministic dynamic principle at the phenomenal level). The result is that (by psychophysical correlation) a definite brain state and state of the observed object are also selected. The same might apply to the connection between consciousness and non-conscious processes in the brain: when superposed non-conscious processes threaten to affect consciousness, there will be some sort of selection. In this way, there is a causal role for consciousness in the physical world. (\cite[pp.262-3]{Chalmers2003})
    \end{quote}

Chalmers in effect combines Wigner’s suggestion that consciousness cannot superpose with Albert’s suggestion that consciousness collapses its physical correlates.  The key idea here is that consciousness is a superposition-resistant property and that its physical correlates therefore resist superposition too. That is, it is difficult or impossible for a subject to be in a superposition of two different states of consciousness, and this results in the collapse of physical processes that interact with consciousness.\footnote{\textcite{Halvorson2011} also argues for a picture on which mental states cannot be superposed and therefore bring about collapse in the physical world.} 

Here the relevant states are total conscious states of a subject at a time.  The total conscious state of a subject is what it is like to be that subject: if what it is like to be subject A is the same as what it is like to be subject B, then A and B are in the same total conscious state. A subject’s total conscious state at a time may include many aspects: visual experience, auditory experience, the experience of thought, and so on.  Like position or mass or color or shape, consciousness in this form can take on many specific values.  Its specific values are the vast range of possible total conscious states of a subject at a time. 

This view assumes that there is a physical correlate of consciousness (PCC): a set of physical states that correlate perfectly with a system's conscious states.  For simplicity, we can start by assuming a materialist view where the total conscious state and its physical correlate are identical.  Things work best if we also assume that the physical correlate of consciousness (PCC) can itself be represented as a quantum observable with an associated operator. This assumption is nontrivial, as not every physical property is an observable; we return to it later.  A PCC observable will have many different eigenstates corresponding to distinct total states of consciousness.  This makes it straightforward to treat consciousness as a super-resistant property. 

To illustrate how this works, we can again suppose an electron in a superposition of locations (again omitting amplitudes for simplicity) $\ket{a} + \ket{b}$.  The electron registers on a measurement device and then the result is perceived by a human subject.  Assuming the measurement device is not conscious, then at the first stage the electron and the device will go into an entangled state $\ket{a}\ket{M(a)} + \ket{b}\ket{M(b)}$.  When the human looks, this result will affect the eye (E), early areas of the nervous system and brain (B), and eventually the physical correlates of consciousness (PCC).  Under Schrödinger evolution, we would expect the electron, device, and subject to go into an entangled state $\ket{a}\ket{M(a)}\ket{E(a)}\ket{B(a)}\ket{PCC(a)}  + \ket{b}\ket{M(b)}\ket{E(b)}\ket{B(b)}\ket{PCC(b)}$.  However, this superposed state would yield a superposition of states of consciousness. So at the point where the PCC is affected, the system will collapse.  It collapses into $\ket{a} \ket{M(a)}\ket{E(a)}\ket{B(a)}\ket{PCC(a)}$ or $\ket{b} \ket{M(b)}\ket{E(b)}\ket{B(b)}\ket{PCC(b)}$, with Born rule probabilities.  In effect, at the point where the measurement reaches consciousness, the electron, the measurement device, and the brain will collapse into a definite state.

On a dualist view on which consciousness merely correlates with physical properties, things are a little more complicated.   We focus on forms of dualism where there are psychophysical laws correlating physical states of a system with states of consciousness.  There will be a set of physical correlates of consciousness (which may be disjunctive if necessary) that are in one-to-one correspondence with total states of consciousness.  A subject will be in a given state of consciousness if and only if it is in the corresponding PCC state.  We can assume as before that the PCC is a quantum observable.   Psychophysical laws connect unsuperposed PCC eigenstates to unsuperposed states of consciousness.   They also connect superpositions of PCC states to the corresponding superpositions of states of consciousness.  A given subject's PCC is in a superposition of PCC eigenstates with certain amplitudes if and only if the subject’s conscious experience is in a superposition of the corresponding total states of consciousness with the same distribution of amplitudes. 

On a dualist view, a fundamental principle will say that consciousness resists superposition.  Whenever Schrödinger evolution plus the psychophysical laws entail that a system enters or is about to enter a superposition of total states of consciousness, the system will collapse into a definite total state of consciousness.  As a result, the PCC will also collapse into an eigenstate, and other physical entities that are entangled with the PCC will collapse as described above.

One motivation for the super-resistance consciousness-collapse model is given by Wigner’s suggestion that superpositions of consciousness are “absurd”.  That is, something about the very nature of consciousness or the concept of consciousness rules out total states where consciousness is superposed. It is certainly at least very hard to imagine subjects who are in superposed states of consciousness (at least without these states becoming total states of consciousness in their own right).  If something about the nature of consciousness explains why it cannot be superposed, then this might provide a possible explanation of why collapse comes about.  This explanatory motivation might be seen as a further motivation for understanding consciousness as the trigger of collapse.

Taking Wigner’s motivation seriously leads to the idea that consciousness is \textit{absolutely superposition-resistant}: that is, that it can never enter superpositions, even brief and unstable ones.  Invoking absolute superposition-resistance leads to a clean and simple dynamic model for collapse involving superselection rules.  Unfortunately this model leads to a fatal problem for absolute super-resistance, which we explore in the next section.

\section{Superselection and the Zeno problem}

To develop super-resistance models in more detail, we can start by thinking of them independently of consciousness.  In principle any observable could serve as a super-resistant observable, with distinct models of quantum mechanics arising from taking different observables to resist superposition.  Later we can consider the special case where consciousness or its physical correlates serve as super-resistant observables.

The simplest (albeit fatally flawed) super-resistance model invokes superselection: the strong form of super-resistance where certain superpositions are ruled out entirely.  In particular, it invokes the familiar concept of a superselection rule: a rule postulating that superpositions of a specified observable are forbidden.  

Superselection rules are invoked for a number of purposes in quantum mechanics.\footnote{Superselection rules were introduced by \textcite{WickWightmanWigner1952}.  There are many somewhat different definitions of superselection rules, analyzed thoroughly by \textcite{Earman2008}.  Here we use a common informal definition.  Superselection rules are invoked in analyses of the measurement process by \textcite{Bub1988}, \textcite{Hepp1972}, \textcite{MachidaNamiki1980}, and others.  \textcite{Thalos1998} gives an excellent review.  The most common strategy is to argue that superselection rules can emerge from the Schrödinger dynamics governing the interaction of a system with its environment.  It is unclear to us whether anyone has explicitly proposed a superselection collapse interpretation, but we are open to pointers.}   Sometimes they are postulated to analyze quantum-mechanical properties that are never found in superpositions, such as the difference in charge between a proton and a neutron. Sometimes they are used to help analyze quantum-mechanical symmetries.  Sometimes they are used to help address measurement in quantum mechanics, most often through the idea that superselection can emerge through interaction with the environment by Schrödinger evolution alone.  

Here we are exploring a somewhat different idea: the idea of a \textit{superselection collapse model}, with a fundamental superselection rule governing the collapse process.  Such a model will specify a superselection observable, such that physical systems must always be in eigenstates of the operator corresponding to the observable. The associated collapse postulate says that whenever a system would otherwise enter a superposition of eigenstates of this operator (given Schrödinger dynamics alone), it instead enters a definite eigenstate, with probabilities given by the Born rule. In the special case where consciousness (or its physical correlate) is a superselection observable, then whenever consciousness would otherwise be about to enter a superposition, it must collapse to a definite state according to the Born probabilities. 

To specify the dynamics better, we can first suppose that the collapse takes place at a time interval of $\Delta t$, so that if the system has evolved (according to the Schrödinger equation) in the preceding $\Delta t$ into a non-eigenstate of the superselection observable, it collapses probabilistically into an eigenstate of that operator, with probabilities given by the Born rule.  This yields a well-defined stochastic process.  For the absolute super-resistance model, the dynamics is the limiting case of this process as $\Delta t$ approaches zero.

The superselection collapse model has a dynamics that is already familiar in quantum mechanics: it is precisely the dynamics that would obtain (on a traditional measurement interpretation) if the resistant observable were being continuously measured by an outside observer. The current approach does not require that there are any outside observers, or that resistant properties themselves are ever measured, or that continuous measurement ever takes place (though to aid the imagination, one could metaphorically suppose that God is continuously measuring the resistant properties of the entire universe). All that it requires is the mathematical dynamics associated with continuous measurement of resistant properties, which is fairly straightforward.

Unfortunately, the dynamics of continuous measurement leads to a well known effect, the quantum Zeno effect, which renders any superselection collapse model empirically inadequate.  The quantum Zeno effect is the effect whereby the more often one measures a quantum observable, the harder it is for the system to enter different states of that observable. In the extreme case where an observable is measured continuously, it cannot change at all.

The source of the quantum Zeno effect lies in the mathematical fact that for a system to evolve under Schrödinger evolution from some initial eigenstate of an operator to some other eigenstate of that operator, it must evolve through superpositions of eigenstates.\footnote{One could argue that this mathematical fact is the common explanation both of the Zeno effect and of the problem for superselection collapse models, rather than the Zeno effect explaining the problem.  Still, the problem is still aptly called a Zeno problem, tied to the impossibility of motion.}  Eigenstates are orthogonal to each other, so the continuous process of Schrodinger evolution cannot evolve directly from eigenstate to eigenstate.  If a system governed by this process cannot pass through superpositions of these eigenstates, then the system cannot change from one eigenstate to another.  Another way to put things is that if small superpositions are permitted, an initial superposition will assign probability 1-$\epsilon$ (where $\epsilon$ is negligible) to the initial eigenstate. So if there is a measurement of this observable in the first moment, the superposition will collapse to the initial eigenstate with probability 1-$\epsilon$. Continuous measurement will therefore force the system to remain in that initial eigenstate.

This leads to the Zeno problem for superselection collapse interpretations.   If there is a superselection observable (one that can never enter superpositions), every system will remain forever in a single eigenstate of that observable.  This consequence may be acceptable for standard superselection observables in physics (such as the charge difference between a proton and a neutron), but it is clearly unacceptable for observables tied to measurement that serve as triggers of the collapse process.\footnote{Mariam \textcite[p.538]{Thalos1998} raises a version of this problem for superselection-based accounts of measurement, arguing that if a classical quantity is governed by a superselection rule, it can never change its magnitude in evolution over time.}   For example, if a superselection observable corresponds to the position of the pointer on a measurement device, then that pointer will be forever stuck in one location and unable to give useful measurement results.

We can illustrate the Zeno problem by taking the superselection observable to be consciousness (or its physical correlate).  We know that systems have different conscious states at different times, and sometimes evolve from being unconscious to being conscious.  If consciousness or its physical correlate was a superselection observable, it would obey the dynamics of continuous measurement so it could not change at all.  If we started in an unconscious state, we could never become conscious.  The unfortunate consequence would be that we could never wake up from a nap.  Furthermore, if there is no consciousness in the early universe, then consciousness could never emerge later.\footnote{Barry \textcite{Loewer2002} raises a different early-universe problem for consciousness-collapse theories: if the first collapse requires the universe to be in a non-null eigenstate of consciousness, then this will never happen, while if collapse is triggered by any superposition of consciousness, then the first collapse will happen too early.  The absolute super-resistance model takes the second horn.  On this view, Loewer’s “collapse too early” problem can be minimized by having conditions for consciousness that are not satisfied in the early universe (so that in its early stages, the universe will be in a null eigenstate of consciousness), and also by noting that most initial collapses when they occur will be onto a null state of consciousness.  The Zeno problem as it arises for the early universe is the distinct but related problem that all collapses will be onto a null state of consciousness.}

The Zeno problem is not just a problem for superselection collapse interpretations.  In “Zeno Goes to Copenhagen”, we argue that the Zeno problem is a serious problem for almost any measurement-collapse interpretation of quantum mechanics.  Any such interpretation faces the question of whether measurement itself can enter quantum superpositions.  If measurement can enter superpositions, the standard dynamics of collapse upon measurement is ill-defined, and new dynamics is required.  If measurement cannot enter superpositions, the quantum Zeno effect suggests that measurements can never start or finish, at least if measurement is an observable.  One way out is to deny that measurement is an observable, but this option leads to further commitments (embracing a strong form of dualism or construing measurement as a special wave-function property) that themselves require a highly revisionary approach.  

In this article, however, we are focusing on the Zeno problem as a problem for super-resistance interpretations.  To handle the Zeno problem in this framework, the obvious move is to abandon  superselection (on which superpositions of the relevant observable are entirely forbidden) for a weaker version of super-resistance. An {\em approximately super-resistant} observable is one that can enter superpositions but nevertheless resists superposition, at least in some circumstances.  On a simple version of this view, superpositions of the observable in question are unstable and they probabilistically tend to collapse over time.

To make the idea of approximate super-resistance precise, we require nonstandard physics. Fortunately, there is a wealth of resources for developing such physics in the literature on modern dynamical collapse theories (\cite{Bassietal2013}).  In section 6, we show how these theories can be adapted to yield a model on which consciousness is approximately super-resistant.  The rough idea is that as a total state of consciousness (and/or its physical correlate) enters increasingly large superpositions (where a large superposition is roughly one that gives significant amplitude to distant states), this yields higher probabilities of collapse of consciousness onto a more definite state. Admittedly it is far from clear what a superposition of states of consciousness would amount to. We return to this matter in the final section.

\section{Integrated information theory}

There are many ways to spell out the details of a consciousness-collapse super-resistance model.  We can combine the view with many different theories of consciousness, and with various different accounts of the collapse dynamics. In what follows we spell out one way of working out some details, by combining the theory with a specific theory of consciousness (integrated information theory, or IIT) and a specific model of approximate super-resistance dynamics (inspired by Pearle’s continuous spontaneous localization interpretation of quantum mechanics).

We focus on IIT for several reasons. First, it is one of the few mathematically precise theories of consciousness. Second, unlike many competitors it purports to be a fundamental theory of consciousness that offers basic and universal principles connecting consciousness to physical processes.  Third, it offers a specific physical correlate for total states of consciousness, using its notion of a Q-shape (qualia shape).  Fourth, it has a distance metric between total states of consciousness, which plays an important role in our framework.  None of this means that we are endorsing IIT.  Many objections have been made to IIT (e.g. \textcite{Aaronson2014}, \textcite{Bayne2018}, \textcite{BarrettMediano2019}, \textcite{Doerig2019}) and they raise important issues.  Our approach could in principle be combined with any theory that has the four properties just listed.  

IIT is a theory that associates systems with both quantitative amounts of consciousness and qualitative states of consciousness.  Its systems are classical Markovian networks made up of interconnected units that interact with each other according to deterministic or probabilistic rules.  Each unit can take on a number of states, and the state of the system is made up of the states of each of the units in the system.

One limitation of IIT as it stands (\textcite{BarrettMediano2019}) is that its assigns amounts and states of consciousness to discrete Markovian network systems but not to real physical systems.  To apply it to real physical systems, we  need to combine it with a mapping from physical systems to network structures.  In what follows we will assume such a mapping (or some other generalization of IIT) so that IIT applies to physical systems. 

IIT is derived from phenomenological axioms rather than from experimental evidence.  Experimental support for it is somewhat limited to date, especially because it is impractical to measure and calculate its measures of consciousness in biological systems. However, some measurable approximations of its quantitative measures have been shown to correlate with level of consciousness, see \textcite{Massiminietal2005}, \textcite{Casarottoetal2016}, \textcite{Leungetal2020}, and \textcite{Afrasiabietal2021}. Additionally, spatiotemporal patterns of integrated information (approximating IIT's qualitative measures) have been derived from brain areas and correlated with the contents of conscious perceptions of faces and other objects (\textcite{Haunetal2017}).  In any case, we will treat IIT as a potential empirical theory of consciousness.  Much of our discussion should generalize to other theories.

IIT is built around the notions of information and integration.  The information in a system is a measure of the extent to which the present state of a system constrains its potential past and future states. One centerpiece of IIT is its measure of integration, which it labels $\Phi$. $\Phi$ is a measure of the extent to which the information in a system is irreducible to the information of its components.  It quantifies how much the causal powers of a system fail to be accounted for by any partitioned version of it. 

The simplest system with nonzero $\Phi$ is a dyad: a network AB with two interacting nodes A and B that swap their states.  If A is on or off, B turns on or off at the next time step, and vice versa.  In this case, AB has causal powers that are not reducible to those of A and B taken alone, and $\Phi$(AB) = 1.  (We  spell out the mathematics in an appendix.)  By contrast, if A and B are not interacting, then the causal powers of AB are reducible to those of A and B taken alone, so $\Phi$(AB) = 0.

IIT says that a system is conscious if and only if it is a maximum of $\Phi$: that is, if the system has higher $\Phi$ than any system nested within it and higher $\Phi$ than any system it is nested within.  The amount of consciousness in a system is $\Phi^{max}$, which is equivalent to $\Phi$ if the system is a maximum and 0 if the system is not.  In what follows we drop the superscript for simplicity.

One way to combine IIT with a super-resistance model is to say that $\Phi$ is super-resistant.  That is, $\Phi$ resists superposition and superpositions of $\Phi$ trigger collapse. Unfortunately, this view faces a fatal problem. It fails to suppress superpositions of qualitatively distinct conscious states with the same value of $\Phi$. Consider a conscious subject and a screen in a dark isolated room. The screen can display green or blue. If it is put into a superposition of displaying both, then the subject will be put into a superposition of experiencing green and experiencing blue. There is no reason to assume that these experiences differ in their $\Phi$–value. But then there is no $\Phi$–superposition, and so no collapse. The subject remains in a superposition of qualitatively distinct total states of consciousness. Such a theory therefore will not yield determinate experiences for many crucial observations.  The underlying problem is that $\Phi$ is not a genuine physical correlate of consciousness – that is, it is not a physical correlate of a total state of consciousness.  It is merely a physical correlate of a scalar degree of consciousness, where the same degree can be present in many different conscious states.\footnote{We canvassed the idea of using $\Phi$ as an absolutely super-resistant property in an early version of this article that raised the Zeno problem for absolute super-resistance and suggested approximate super-resistance via continuous localization as a possible solution. In an article responding to our early presentation and building on the ideas there, \textcite{OkonSebastian2018} develop the idea that $\Phi$ could be an approximately super-resistant property using continuous localization. Okon and Sebastian respond to our current objection by saying that decoherence makes it extremely unlikely that there will be superposed conscious states with the same value of $\Phi$. The blue/green case seems a clear case of this sort of superposition, however, as does any ensuing state resulting from interactions with their environment that makes no difference to their total state of consciousness. The dyad system discussed in the main text and the appendix gives a simple illustration of a superposition of states with different Q-shapes but with the same value of $\Phi$.  In addition, the Q-shape collapse model is much better suited for giving all aspects of consciousness a causal role, whereas the $\Phi$-collapse model gives degree of consciousness a causal role and leaves everything else epiphenomenal.}

Fortunately, IIT also postulates a physical correlate of total states of consciousness.  The Q-shape (qualia shape) of a system is an entity that serves as an abstract representation of the structure of the integrated information in a system. IIT specifies a mathematical mapping from network structures to Q-shapes.  If we assume (as above) that the total physical state of a system determines a network structure, then IIT will derivatively specify a mapping from total physical states to Q-shapes.

The Q-shape of $S$ is a set of  weighted points, one for each mechanism in $S$. A mechanism is a subsystem $m$ of $S$ -- that is, a nonempty set of elements of $S$ -- with $\phi(m)>0$ (as defined in Appendix A).   If $S$ has $n$ elements, then it has up to $N=2^n-1$ mechanisms.  For example, in the dyad system AB, which has two elements A and B, the subsystems are A, B, and AB, and the mechanisms are A and B.  The weight associated with a mechanism $m$ is $\phi(m)$, a non-negative real number representing the integrated information associated with $m$.  The point associated with $m$ is given by two probability distributions over the $2^n$ states of $S$, the so-called maximally irreducible cause repertoire and maximally irreducible effect repertoire associated with $m$. 

According to IIT, a system’s Q-shape determines (at least nomologically) the total state of consciousness associated with that system.  A Q-shape is itself a mathematical entity, and it is not obvious just how a Q-shape determines a state of consciousness.  What matters most for our purposes is that according to IIT, (i) having a given Q-shape is a physically definable property (we might call it physical Q-shape), (ii) Q-shape is a physical correlate of consciousness, in that any two physical systems with the same associated physical Q-shape will have the same state of consciousness. It will also be helpful to assume the stronger theses that (iii) the mathematical structure of a conscious state is given by a Q-shape (call this a system's phenomenal Q-shape) and (iv) as a matter of psychophysical law, a system has a given phenomenal Q-shape (that is, it has a conscious experience with a given structure) if and only if has the isomorphic physical Q-shape (that is, it has a physical state with the same structure as defined by IIT).  These claims are far from obviously correct, but something like them seems to be intended by IIT.

As before, it does not matter too much for our purposes whether these claims of IIT are correct.  It is plausible that a final mathematical theory of consciousness will specify \textit{some} mathematical structure for consciousness (though there may be more to consciousness than its mathematical structure, as inverted qualia cases suggest).  And it is plausible that this mathematical structure should be realized in some way in the physical correlates of consciousness.  If necessary, we can replace Q-shape by that mathematical structure.  What matters most is that there is some precise theory of consciousness for which psychophysical isomorphism principles like this are correct.

Different states of the dyad system AB discussed earlier can be associated with different Q-shapes.  Consider state 10, where A is on and B is off, and state 00, where both A and B are off.  As we show in the appendix, both states have $\Phi=1$, but they are associated with distinct Q-shapes.  In principle one can prepare a dyad system in a superposition of these two states 10 and 00: we might call this \textit{Schrödinger’s dyad}.  If Q-shape is super-resistant, Schrödinger’s dyad will be unstable and will collapse into a state with a definite Q-shape.  We discuss a framework for combining IIT with quantum mechanics along these lines in the next two sections.  In section 7, we discuss possible experimental tests, which are likely to rule out the simple Q-shape collapse interpretation but which suggest a program for empirically refining collapse interpretations.

  \section[Combining IIT with quantum mechanics]{Combining IIT with quantum mechanics\footnote{This section is co-authored with Johannes Kleiner (Münich Center for Mathematical Philosophy, Ludwig Maximilian University).}}

The standard IIT framework (\textcite{Oizumietal2014}) maps classical network states to Q-shapes.  We have assumed a derivative mapping from classical physical states to Q-shapes.  To combine IIT with quantum mechanics, we need to extend the IIT mapping so that it maps quantum physical states to Q-shapes or to superpositions of Q-shapes.  The core idea of a Q-shape collapse model is that systems in superpositions of Q-shape always collapse toward having a determinate Q-shape.

To extend the IIT mapping to quantum physical states, the obvious way to proceed is to use IIT's physical definition of Q-shape to define a set of Q-shape collapse operators, one for each dimension of Q-shape. The joint eigenstates of these operators will be physical states with determinate Q-shapes.

A challenge to defining these Q-shape operators is that in the classical IIT framework, $\phi$ and Q-shape depend on probabilities of  state-transitions in a network, which may depend on the position and
momentum of the system’s parts. Position and momentum are noncommuting operators, so physical systems cannot be in joint eigenstates of them. High-mass systems may have precise enough position and momentum to determine $\phi$ and Q-shape, but these quantities may not be defined for low-mass entities such as electrons in quantum systems (\textcite[p97]{McQueen2019a}).

There are various options for addressing this challenge.  We could redefine $\Phi$ and Q-shape so they depend only on positions or mass densities of elements of the system.  We could also give special treatment for low-mass systems, for example modifying $\Phi$ to stipulate that $\Phi=0$ for systems with mass below a certain threshold, or we could invoke a coarse-grained or ``smeared" version of $\Phi$ and Q-shape observables, with significant smearing mainly required for systems with very low mass.

Alternatively, we can invoke newer versions of IIT that are defined over quantum states.   One framework for an IIT-driven collapse model has been developed by \textcite{KremnizerRanchin2015}, who define a new measure of quantum integrated information QII for quantum systems.  On their model, a system's QII determines the probability of collapses onto a position basis, so that systems with higher QII are more likely to collapse on to the position basis. However, Kremnizer and Ranchin’s interesting model is a super-resistance theory only in a weak sense: the properties that trigger collapse (QII) are quite distinct from the collapse basis (position), and position resists superposition only in certain contexts with high QII.  Also, while Kremnizer and Ranchin speculate that their quantity QII may be a measure of consciousness, this will yield at best a  limited causal role for consciousness, on which the scalar amount of consciousness determines probability of collapse but the specific conscious state of a subject plays no role. 

\textcite{zanardi2018quantum} have developed a more thoroughgoing quantum-mechanical version of IIT, defining quantum mechanical operators for each IIT notion (including Q-shape as well as $\phi$) across a broad class of quantum-mechanical networks.  (These are networks of finite-dimensional non-relativistic qudits, interacting via Markovian trace preserving completely positive maps.) Further generalizations have been given by \textcite{KleinerTull2020}.  These models do not yet give a complete mapping from physical states to Q-shapes, but they come closer to doing this than standard IIT.  In what follows, we will assume a fully developed model along these lines with a complete mapping from physical states to Q-shapes.

Quantum IIT specifies a mapping $E$ from states of quantum systems to {\em quantum Q-shapes}.  Quantum Q-shapes are
are quantum analogs of  classical Q-shapes, the Q-shapes invoked in standard IIT.     Classical Q-shapes for an $n$-element system $S$ can be represented as $N=2^n-1$ weighted points, one for each subsystem of $S$, where points are pairs of probability distributions and weights are non-negative real numbers (for a subsystem that is not a mechanism, the weight will be zero).  Quantum Q-shapes likewise involve $N$ weighted points, where points are now pairs of density operators associated with the Hilbert space of $S$ and weights are non-negative reals.  Where the space of classical Q-shapes is the Cartesian product of $N$ copies (one for each subsystem) of $Pr(S) \times Pr(S) \times R^+_0$, the space of quantum Q-shapes is the Cartesian product of $N$ copies of $D(S) \times D(S) \times R^+_0$.  Here $Pr(S)$ is the space of probability distributions over $S$, whose quantum analog $D(S)$ is the space of density operators over $S$.  $R^+_0$ is the set of non-negative real numbers.\footnote{If $S$ is a network of elements with binary states, each weighted point will have $2^{n+1}+1$ dimensions (two $2^n$-dimensional probability spaces plus a real number), so classical Q-space has $(2^n-1)(2^{n+1}+1)$ dimensions.  In quantum IIT, the $2^n$ dimensional probability-spaces are replaced by $2^{2n}$-dimensional density spaces, so quantum Q-space has $(2^n-1)(2^{2n+1}+1)$ dimensions.} 

There is a natural mapping from classical Q-shapes to a subclass of quantum Q-shapes,  deriving from a mapping from $Pr(S)$ to $D(S)$, defined as follows: $(p(s_i)) \mapsto \sum_i p(s_i) \ket{s_i} \bra{s_i}$.  We can call this distinguished subclass of quantum Q-shapes the {\em quasi-classical} Q-shapes. 
Any quantum Q-shape can be seen as a superposition of quasi-classical Q-shapes.

Quantum IIT as it stands does not say much about how quantum Q-shapes correspond to states of consciousness.  For our purposes we can add the further claims that (i) quasi-classical Q-shapes correspond to determinate states of consciousness, exactly as the corresponding classical Q-shapes do in classical IIT, and (ii) other quantum Q-shapes are superpositions of quasi-classical Q-shapes and correspond to superpositions of the corresponding states of consciousness.

We can define the quasi-classical {\em states} of a quantum system as those quantum states that quantum IIT associates (via the mapping $E$) with a quasi-classical Q-shape.  If $\mathcal C$ is the class of quasi-classical Q-shapes, the class of quasi-classical quantum states is $\mathds{E}^{-1}(\mathcal C)$, the preimage of $\mathcal C$ under $\mathds E$.  Every state of a quantum system can then be represented as a superposition of quasi-classical states, and its associated Q-shape will be a superposition of the corresponding quasi-classical Q-shapes.  We can then set up collapse operators so that quantum systems always collapse toward these quasi-classical states with quasi-classical Q-shapes.

One limitation of quantum IIT as it currently stands is that these quasi-classical states (picked out as those that quantum IIT associates with quasi-classical Q-shapes) may not closely correspond to what we usually think of as quasi-classical quantum states such as mass density eigenstates.  As a result, the Q-shape collapse dynamics need not lead to collapse toward standard ``classical" states such as mass density eigenstates and may result in a superposition of these states (along with a relatively determinate state of consciousness).  If we want to avoid these quantum superpositions as physical correlates of determinate consciousness, there is at least a research program of developing a version of quantum IIT on which quasi-classical Q-shapes and determinate states of consciousness are associated with more ``classical" quantum states. In what follows it may be helpful to assume such a version of the framework.  

We can now define Q-shape collapse operators.  Recall that a Q-shape is a point in the direct product of $N$ copies of the density operator space $D(S)$.  Any density operator in $D(S)$ can be represented (in the quasi-classical basis $\ket{s_i}$) as 

\begin{equation}
    \rho =  \sum_{i,j} c_{ij} \ket{s_i}\bra{s_j}
\end{equation}

The Q-shape for any given quantum state $\psi$ consists of $2N$ density operators of this kind and $N$ non-negative real numbers. The Q-shape can therefore be  represented by $2N$ sets of coefficients $c_{ij}^k$ which we denote as $c_{ij}^k(\psi)$ (for $k=1 \ldots 2N$), and $N$ non-negative real numbers which we denote $\varphi^k(\psi)$ (for $k=1\ldots N$). For notational simplicity, we duplicate each of the latter, so that for each $k=1, \, ... \, 2N$, we have a $c_{ij}^k(\psi)$ which describes the first or second factor in $D(S) \times D(S) \times R^+_0$ and a $\varphi^k(\psi)$ which describes the third factor.

We can then define an ensemble of orthogonal self-adjoint collapse operators as follows:

   \begin{equation}\label{Kleiner}
            \hat Q^k_{ij} := \sum_{\psi \in \mathds{E}^{-1}(\mathcal C)}
    \varphi^k(\psi) ((c_{ij}(\psi)+(c_{ji}(\psi)) \ket{\psi} \bra{\psi}.
    \end{equation}
    
  The sum has been restricted so that it runs over the class $\mathds{E}^{-1}(\mathcal C)$ of quasi-classical quantum states, that is, those whose Q-shapes are quasi-classical.  As $k$ ranges from 1 to $2N$ (where $N=2^n-1$) and $i$ and $j$ range from 1 to $2^n$, an $n$-element system will be associated with $2^{2n+1}(2^n-1)$ collapse operators.\footnote{For reasons tied to the role of weights within IIT, we have combined the real weights $\varphi^k$ and the density operator coefficients $c_{ij}$ in defining the Q-shape collapse operators.  As a result there are $N$ fewer collapse operators than dimensions of Q-space.  Alternatively one can define separate operators for the coefficients and real weights as follows: $\hat Q^k_{ij} := \sum_{\psi \in \mathds{E}^{-1}(\mathcal C)}
   ((c_{ij}(\psi)+(c_{ji}(\psi)) \ket{\psi} \bra{\psi}$ and $\hat B^l := \sum_{\psi \in \mathds{E}^{-1}(\mathcal C)}\varphi^l(\psi) \ket{\psi} \bra{\psi}$.}

\section{Continuous collapse dynamics}

To complete our picture of super-resistant consciousness-based collapse, we need an account of the dynamics of super-resistant collapse.  Fortunately, there exist models of dynamic collapse (due to Philip Pearle and Lajos Di\'osi, among others) that can be generalized to model the continuous collapse of any observable.  It is not difficult to adapt these models to model the continuous collapse of consciousness and its physical correlates such as Q-shape.  We start by informally reviewing these models and the adaptation to consciousness-collapse models, before providing formal details.\footnote{Thanks to Maaneli Derakhshani, Philip Pearle, and  Johannes Kleiner for their extensive help with the material in this section.}

We start with the continuous spontaneous localization (CSL) model due to Pearle (\cite*{Pearle1976}, \cite*{Pearle1999}, \cite*{Pearle2021}). Pearle's model is a continuous relative of the well known GRW model, on which the position of isolated particles undergo spontaneous localization of position with low probability at any given time.  On CSL, wave functions undergo a gradual stochastic collapse process at all times.  The model provides continuous collapse onto mass density: the amount of mass present at various locations.  It provides a dynamics by which superpositions of mass density gradually collapse toward definite states of mass density, with faster collapse in high-mass systems.  In effect, CSL is a model on which mass density is super-resistant.

Pearle’s model can be informally motivated by an analogy between gradual collapse and the gambler's ruin game in classical probability theory (\cite{Pearle1982}). In the gambler's ruin, a number of gamblers play against each other until all but one of them is “wiped out”.  Consider two gamblers, $G_1$ and $G_2$, who have \$100 between them such that $G_1$ has \$60 and $G_2$ has \$40. They toss a coin: if heads $G_1$ gives a dollar to $G_2$, if tails $G_2$ gives a dollar to $G_1$. As they keep playing, their respective amounts fluctuate, but the total remains the same. Eventually, the game ends, as one player acquires \$100. It turns out that $G_1$ wins 60\% of the time while $G_2$ wins 40\% of the time. That is, the probability that a given gambler wins is determined by the initial stakes.

In CSL, the squared amplitudes in a superposition (in the preferred basis) play a continuous stochastic gambler's ruin game against each other, fluctuating up and down until one “wins”, thereby completing the collapse. The probability that a given state vector “wins” a collapse in the long run is determined by its initial squared amplitude according to the Born rule. Crucially, we may control the speed at which the games are played in terms of certain (experimentally bounded) parameters. This allows large superpositions to collapse quickly and small superpositions to collapse at a negligible rate.

Like the GRW theory, Pearle’s theory involves a weak sort of super-resistance.  Mass density resists superposition weakly, in that an isolated particle will only gradually collapse toward a definite position and so a definite mass density.  At the fundamental level, superpositions of mass density will be ubiquitous.  However, when many particles are entangled in a macroscopic system, the mass density of the system as a whole will collapse extremely fast, so that we will never encounter macroscopic systems in large superpositions of mass density.

Continuous collapse models can be adapted to work with super-resistant properties other than position and mass density.  Given any observable, we can postulate a continuous collapse process with a version of the Pearle dynamics applied to this observable.  Squared amplitudes for eigenstates of the observable engage in a stochastic gambler's ruin process, so that systems in superpositions of the observable collapse quickly or slowly toward their eigenstates via a gamblers-ruin process.

A related collapse process is postulated in the \textcite{Penrose2014} model of gravitational collapse, where spacetime curvature is super-resistant.  Superpositions of spacetime curvature collapse onto definite states.  Unlike Pearle, Penrose does not give a fully defined dynamics for collapse.  He defines a superposition lifetime, $\hbar/\Delta E_G$, where $\hbar$ is Planck’s constant and $\Delta E_G$ is the gravitational self-energy of the difference between the mass distributions belonging to the two states in the superposition. But the dynamics of collapse during this lifetime are not specified.\footnote{The \textcite{HameroffPenrose} ``Orch OR" model extends Penrose's model of collapse into a model of consciousness.  The Penrose-Hameroff model is not a consciousness-collapse model either: Penrose and Hameroff hold that collapse is triggered by superpositions of spacetime curvature rather than by consciousness or measurement, and that collapse causes consciousness rather than vice versa.  Our approach might be considered a distant cousin of the Penrose-Hameroff model, with the main differences on our approach being: (i) consciousness causes collapse rather than vice versa, (ii) collapse is onto Q-shape rather than onto spacetime curvature, (iii) the collapse dynamics corresponds somewhat more closely to Pearle's model rather than Di\'osi-Penrose's, and (iv) as discussed later, we make no claims about quantum coherence and quantum computation in the brain.}

An account of the dynamics of gravitational collapse has been independently provided by Lajos \textcite{Diosi1987}.  Di\'osi sets out a stochastic version of the Schrodinger equation on which there is a continuous collapse process onto spacetime structure.  Di\'osi's dynamic collapse process is closely related to Pearle's continuous spontaneous localization process, with some differences arising from the use of a collapse  onto gravitational structure as opposed to mass density.

It turns out that the Di\'osi and Pearle dynamics are both instances of a general formulation of continuous collapse dynamics which can be applied to any collapse operator.  Such a formulation has been presented by Angelo Bassi and coauthors (2017).\footnote{See also \textcite[eqn.10]{Pearle1999} and \textcite[eqn.14]{Bassietal2013}.}  We will adapt this formulation to set out a dynamics for continuous collapse onto consciousness.

In the context of IIT, we can use this general dynamics to develop a view on which Q-shape is super-resistant.  Informally: Suppose a system is in a superposition of two Q-shapes, each with an associated amplitude. We can stipulate a ``localization" dynamics for this superposition that works much like Pearle’s except that collapse is toward eigenstates of Q-shape.  The amplitudes trade off probabilistically with each other over time, in effect playing gambler's ruin at a rate proportional to the distance between the two Q-shapes.  In the long run, the system will collapse onto a specific Q-shape with probability given by its initial squared amplitude.

We can spell out the mathematical details as follows.  The general framework for continuous collapse rests on using a modified version of the Schrodinger equation that includes a nonlinear and stochastic term for collapse as well as the standard linear deterministic evolution.  To be consistent and compatible with constraints such as no superluminal signalling, nonlinear modifications to the Schrodinger equation must take a highly constrained stochastic form.  This yields the following general form for continuous collapse models (\cite[p.27]{Bassietal2017}):

\begin{equation}\label{Bassi(74)}
d\psi_t = [-i\hat{H}_0dt + \sqrt{\lambda}(\hat{A} - \langle\hat{A}\rangle_t)dW_t - \frac{\lambda}{2}(\hat{A}-\langle\hat{A}\rangle_t)^2dt ]\psi_t
\end{equation}

Here $\psi_t$ is the wave function state at $t$, $\hat{H}_0$ is the Hamiltonian, $\lambda$ is a real-valued parameter governing collapse rate, $\hat{A}$ is a collapse operator, $\langle\hat{A}\rangle_t$ is its expected value at $t$, and $W_t$ is a noise function allowing for stochastic behavior.  The equation allows continuous stochastic collapse toward an eigenstate of the operator $\hat{A}$ at a rate governed by $\lambda$ and $W$, with probabilities given by the Born rule.

It is straightforward to generalize this equation to multiple collapse operators.\footnote{\textcite[eqn.36]{Bassietal2013}} Using our Q-shape collapse operators defined in (\ref{Kleiner}), we can propose the following dynamics:

\begin{equation}\label{ourproposal1}
\begin{split}
d\psi_t = [-\frac{i}{\hbar}\hat{H}dt + \sqrt{\lambda}\sum_{\alpha}(\hat{Q}_{\alpha} - \langle\hat{Q}_{\alpha}\rangle_t)dW_{\alpha,t} \\  -\frac{\lambda}{2}\sum_{\alpha}(\hat{Q}_{\alpha} - \langle\hat{Q}_{\alpha}\rangle_t)^2dt]\psi_t.
 \end{split}
\end{equation}

Here $\alpha = i,j,k$ is a multi-index that comprises the indices in (\ref{Kleiner}). If there is little difference in the superposed Q-shapes, then the first term on the right hand side (representing Schrödinger evolution) dominates. Otherwise, the system collapses toward a joint eigenstate of the collapse operators, at a rate proportional to the sum of the difference between their eigenvalues.

The noise function $W_{\alpha,t}$ is responsible for the stochastic ``gambler's ruin" collapse behavior described earlier.\footnote{For a simple illustration of how this works, see \textcite[sec. 2.2]{Pearle1999}.}  In CSL and other mass density collapse models, the collapse operators correspond to local mass densities $\hat{m}(x)$ (the amount of mass at location $x$).  The CSL noise function is given by Wiener processes $W_t(x)$, representing Brownian motion through time at location $x$.  The noise at different spatial locations $x$ and $y$ is correlated by a spatial correlation function $G(x-y)$, which in CSL is a Gaussian function of the distance between $x$ and $y$.  This ensures that collapse rate depends on the distance between mass density distributions.

In our IIT-based collapse model, we can define the collapse rate so that it depends on the extended Earth movers distance EMD* between Q-shapes (see the appendix). To ensure this, we can stipulate that the spatial correlation function involved in the noise functions $W_t(x)$ is defined in terms of Earth movers' distance: specifically, $G(x,y)=1/EMD^*(x,y)$ (with an appropriate cut-off for when $EMD^*(x,y)$ is small or zero; we omit the details).

In CSL it is also standard to ``smear" the mass density operator with the same Gaussian $G(x-y)$, so that collapse is onto smeared mass density eigenstates rather than precise mass density eigenstates, thereby avoiding large violations of energy conservation.\footnote{Smearing the mass density operator results in the following equation: 
$d\psi_t = [-\frac{i}{\hbar}\hat{H}dt + \sqrt{\lambda}\int d^3x(\hat{m}(\textbf{x} - \langle\hat{m}(\textbf{x})\rangle_t)dW_t(\textbf{x})  -\frac{\lambda}{2}\int d^3x \int d^3y  \boldsymbol{G}(\textbf{x - y})(\hat{m}(\textbf{x}) - \langle\hat{m}(\textbf{x})\rangle_t)(\hat{m}(\textbf{y}) - \langle\hat{m}(\textbf{y})\rangle_t)dt]\psi_t$.} Our equation is simpler because our collapse operators do not correspond to points (or smeared regions) in a continuous space but instead correspond to a discrete set of mechanisms. We therefore do not need to include a smearing function in our equation. In principle, however, it is straightforward to add such a smearing function as in mass density models.

Our equation (\ref{ourproposal1}) assumes that all superposed Q-shapes are Q-shapes of a single system (network of units) with a fixed number of units and a fixed causal structure.   It does not address the case where we have a superposition involving Q-shapes of systems with different numbers of units or causal structures.  Extending the current framework to handle those cases is a further project.

The overall theory may look complex, but the underlying principles are fairly simple.  First, there is an IIT-style quasi-classical psychophysical theory linking physical Q-shape by a structural isomorphism to phenomenal Q-shape in states of consciousness.  Second, there is a generalization of this theory to the quantum realm, so that superpositions of physical Q-shape are linked to superpositions of phenomenal Q-shape and so to superpositions of states of consciousness.  Third, there is the key claim that consciousness is super-resistant.  More specifically, phenomenal Q-shapes resist superposition via a Pearle-style principle of continuous collapse for Q-shapes, so that superpositions of consciousness rapidly become more determinate.  Putting these elements together: superpositions in the environment lead to superpositions of Q-shape in the brain, which lead to superpositions of consciousness. These superpositions of consciousness will rapidly collapse, yielding collapse in the correlated Q-shapes and collapse in the brain states and the environmental states that are entangled with Q-shape.

\section{Experimental tests}\label{experiments}

Different super-resistant collapse models make different predictions.  For any proposed super-resistant property, in principle it is possible (though usually extremely difficult) to test whether a system is in a superposition of that property. This means that in principle (although not yet in practice) it is possible to test which systems can collapse quantum wave functions, and in virtue of which of their properties.  For example, in principle we can test whether atoms, molecules, cells, worms, mice, dogs, or humans, as well as oscilloscopes, computers, and other devices have the capacity to collapse a wave function.\footnote{It is occasionally suggested that we know from existing results that ordinary measuring devices collapse the wave function, perhaps because we always find them in definite states, or because their measurements do not lead to quantum interference. However, it is easy to see that these observations are all equally consistent with a view on which only humans (say) collapse wave functions, and measurement devices are observed by humans and entangled with their environment.  Sophisticated variants of this objection are made by \textcite{KochHepp2006} and \textcite{CarpenterAnderson}. \textcite{OkonSebastian2016} explain what goes wrong in these objections.}  

To test whether a given property supports superpositions, one can use an interferometer for this property, which detects interference between superposed quantities in much the same way that a double-slit experiment detects interference between superposed positions. In practice it is extraordinarily difficult to set up interferometers for complex properties instantiated by complex systems, because of the need to prepare the relevant system in complete isolation from environmental effects. To date, the most complex such measurements have detected interference in large molecules with around 2000 atoms (\cite{Feinetal2019}). Current limitations are practical rather than principled, and measurements for more complex properties are certainly possible in principle.

These tests have clear implications for super-resistance models.  In absolute super-resistance models, superpositions of super-resistant observables are impossible.  In approximate super-resistance models, these superpositions are unstable.  So at least on a first approximation: if we detect widespread superpositions of an observable, that tends to disconfirm models on which that observable is super-resistant.

On a second approximation, all this depends on just how unstable the superpositions are.  We can distinguish \textit{fast-collapse} models on which large superpositions of a super-resistant observable are rare, from \textit{slow-collapse} models on which large superpositions are common.  Here a large superposition of an observable is a superposition of significantly different eigenstates of the observable with significant amplitudes for significant periods (where ``significant" is a placeholder for now).   If we frequently detect large superpositions of an observable, this tends to disconfirm at least fast-collapse super-resistance models involving that observable.  These results do not disconfirm slow-collapse models as easily.  Still, where consciousness-collapse models are concerned, fast-collapse models are arguably preferable to slow-collapse models, as the latter allow that large superpositions of conscious states are common.  So for now, we will focus on fast-collapse models, returning to slow-collapse models shortly.

We may already be in a position to test fast-collapse models in which Q-shape is super-resistant. This project is aided by the fact that even quite simple systems (such as a dyad) can have nonzero $\Phi$ and nontrivial Q-shapes, as we have seen.  To test the hypothesis, we need only prepare a quantum computer to enter superpositions of Q-shape.  The simplest example is Schrödinger’s dyad (from section 4): two units A and B in a superposition of connected and disconnected states with distinct Q-shapes. If we find the interference effects predicted by standard quantum mechanics (which assumes that simple systems do not perform measurements and evolve according to Schrödinger dynamics), this will falsify the hypothesis that Q-shape is super-resistant, at least on a fast-collapse model. If we do not find these effects, this will suggest that these superpositions are impossible or unstable and will tend to support the hypothesis that Q-shape is super-resistant.

Something along these lines could be done with a quantum version of a Fredkin crossover gate.\footnote{Thanks to Scott Aaronson for this suggestion.} A classical Fredkin gate involves three bits, a control bit and two other bits A and B.  If the control bit is 1, bits A and B are swapped.  If the control bit is 0, bits A and B are left as is.  In a quantum version of the Fredkin gate, the control bit can be in superposition, and the AB system will then be in a superposition of bit-swapping and staying constant. As a result, IIT appears to suggest that the AB system will be in a Q-shape superposition. If Q-shape is super-resistant in a fast-collapse model, we should expect this superposition to collapse.

In fact, a quantum Fredkin gate has recently been constructed (\cite{Pateletal2016}), with results indicating a successful superposition. However, in this example, it does not seem that the conditions for $\Phi$(AB)=1 are met, because there is no two-way feedback interaction between gates A and B. In IIT, purely \textit{feedforward} networks typically have zero $\Phi$. A feedforward network can have nonzero $\Phi$ if it has overlapping inputs and overlapping outputs, but this does not appear to be happening in the quantum Fredkin gate.\footnote{This points to another test case that can be realized by a quantum computer. Perhaps the simplest feedforward system with nonzero $\Phi$ is a dyad system CD that forms a layer of a feedforward network, whereby a node from a previous layer gives input to both C and D, and both C and D give input to a node in a subsequent layer. For illustration, see \textcite[Fig. 7(B)]{Oizumietal2014}.} 

How might we properly construct feedback systems such as AB using quantum computers?  In the quantum computing literature, two primary types of quantum feedback are distinguished. The traditional type is \textit{measurement-based feedback}. Here, a quantum system performs some (usually feedforward) processing and is measured, and the measurement result is then fed back into the quantum system as input. This will not help for our purposes. A more recent development is \textit{coherent quantum feedback} (\cite{Lloyd2000}), where feedback connectivity obtains in the quantum system itself. Superpositions of coherent quantum feedback could be used to build our dyad system in a superposition of states. 

For example, consider the ion-trap example discussed by \textcite[p4]{Lloyd2000}. The initial state of the system is $\ket{\psi}_s\ket{0}_m\ket{\phi}_c$, where $\ket{\psi}_s$ is the unknown state of the ``system" ion, $\ket{\phi}_c$ is the prepared state of the ``controller" ion and $\ket{0}_m$ is the vibrational mode cooled to its ground state. Lloyd explains how certain directed pulses can evolve the system from $\ket{\psi}_s\ket{0}_m\ket{\phi}_c$ to $\ket{\downarrow}_s\ket{\psi'}_m\ket{\phi}_c$, to $\ket{\downarrow}_s\ket{\phi'}_m\ket{\psi}_c$, and finally to $\ket{\phi}_s\ket{0}_m\ket{\psi}_c$. In effect, the initial unknown state of the system ion is swapped with the initial state of the controller ion.
Schrödinger’s dyad may then be constructed by putting the input pulses into a superposition of implementing this swap and not implementing this swap, yielding: 
$\alpha\ket{\psi}_s\ket{0}_m\ket{\phi}_c + \beta\ket{\phi}_s\ket{0}_m\ket{\psi}_c$.  If the two terms in the superposition yield distinct Q-shapes, then our model predicts that this superposition is unstable and will eventually collapse, even if the system remains isolated.

The issue is not entirely straightforward, as it might be denied that the full conditions for $\Phi$(AB)=1 are met (perhaps because of the role of the vibrational mode or the pulses). Still, it seems likely that some technologically feasible quantum computation involves a superposition of Q-shapes. If found, such a superposition will falsify the combination of standard IIT (on which Q-shape is the physical correlate of consciousness) and the fast-collapse consciousness-collapse thesis.

More generally, most proponents of quantum computing predict that superposed states in larger and larger systems will gradually be demonstrated. It would be foolhardy to bet against these predictions.  In the face of these results, one could maintain an IIT-collapse view by modifying IIT somewhat: for example to say that a system is conscious (and has a Q-shape) only when $\Phi$ is above a certain threshold, or by adding other constraints to the definition of $\Phi$ so that the relevant simple systems have $\Phi$ = 0.  Alternatively one could adopt a slow-collapse version of the model; one could reject IIT entirely for a different theory of consciousness; or one could reject the consciousness-collapse thesis.  Still, this shows how even near-term experimental results from quantum mechanics can have some bearing on theories of consciousness. 

All this brings out that the consciousness-collapse thesis in its fast-collapse version is not easy to combine with panpsychist theories of consciousness on which consciousness is found even in very simple systems.   A strong panpsychist fast-collapse view on which position or mass or charge quickly collapses the wave function is straightforwardly refuted by standard experimental results showing interference effects. The more recent results of Fein et al demonstrating superpositions of position in 2000-atom systems tend to suggest that the threshold for collapse lies somewhere beyond that level. There are some quasi-panpsychist collapse views involving slightly more complex properties distinct from position that have not yet been tested, but we should easily enough be able to test them as above, and few would expect them to be supported.  The consciousness-collapse thesis (in fast-collapse versions) tends to fit more comfortably with non-panpsychist views on which consciousness arises only in relatively complex systems.  These views are consistent with existing and likely near-term-future observations, while still being subject to experimental test eventually. 

There remains the possibility of slow-collapse models on which superpositions of consciousness tend to collapse slowly across long periods.  If these models allow widespread large superpositions of human states of consciousness, these views are hard to reconcile with introspection, and it also becomes less clear why we should accept the consciousness-collapse view over an Everett-style view where one's consciousness is constantly in large superpositions.  Perhaps there could be a CSL-style slow-collapse panpsychist model on which superpositions of consciousness are common but unstable at the microphysical level, in the way that superpositions of mass distribution are common but unstable at the microphysical level in CSL.  In CSL, large superpositions of macroscopic mass distributions are nevertheless uncommon. Likewise, a panpsychist slow-collapse view might have the consequence that large superpositions of human consciousness are uncommon, especially on a constitutive panpsychist view on which human consciousness is constituted by patterns of microconsciousness.  Such a view will face the notorious combination problem of how this constitution works, and it may also have less of an irreducible causal role for human consciousness than other collapse views.  Still, there are various versions of a slow-collapse model worth exploring.

There are also empirical constraints on super-resistance models tied to energy conservation (collapses tend to produce excess energy, so they cannot be too frequent or too dramatic\footnote{The main difficulty in the experimental detection of such effects involves controlling all the possible ways of cooling. Thus, in their discussion of testing GRW and CSL, \textcite{FeldmannTumulka2012} consider the Kubacher Kristallhöhle, the largest natural cave in Germany, which is 9$^{\circ}$ C all year around. When surface temperatures are low, heat spontaneously created in the cave cannot be transported away, thereby suggesting a way of obtaining an empirical bound on the rate of spontaneous warming. It is much more difficult to see how we could find empirical bounds on spontaneous warming in conscious systems, but it may not be impossible.}) and to the quantum Zeno effect (a super-resistance model must allow superpositions to persist long enough to avoid Zeno effects, while not persisting so long that measurements do not have definite outcomes).  All these phenomena impose constraints that narrow the class of available super-resistance models: super-resistant properties are not too simple and not too complex, while collapses are not too frequent and not too slow.

For a super-resistance model to be empirically supported, we will eventually have to find systems and properties that resist superposition. One key (if currently far-fetched) experiment would use an interferometer on a human isolated from their environment, preparing them to enter a superposition of conscious states and seeing if interference effects are observed.  If interference effects are not observed, one will have experimental support for the claim that humans can collapse wave functions.  As before this would not decisively demonstrate that consciousness is doing the work, but it would give reason to take that view seriously.  If interference effects are not observed, one will have experimental support for the claim that humans cannot collapse wave functions.  This will also tend to falsify any measurement-collapse formulation of quantum mechanics, and in particular will tend to falsify the view that consciousness collapses the wave function.  In this way the framework of this article may ultimately be subject to empirical test. 

Admittedly, it is not clear that it will ever be possible to isolate and test a conscious human brain in this way.  Perhaps somewhat more feasible in the long term could be running a detailed simulation of a human brain on a quantum computer.  If interference effects are not observed, one will have experimental support for the claim that the computational structure of the human brain can collapse wave functions.  If they are not observed, one will have evidence against this claim.  However, this result will leave open the hypothesis that other features of the human brain that are not replicated in a simulation, such as biological features, are responsible for wave-function collapse. It may be especially difficult to test biological collapse models, as many standard methods of isolating systems to test for superposition require low temperatures where the biology may break down.  Still, these quantum computing experiments might at least give us evidence for or against a consciousness-collapse model where the correlates of consciousness are computational.  In the long run, advances in quantum computing are likely to heavily constrain the prospects for consciousness-collapse models.

\section{The causal role of consciousness}

On the picture we have sketched, superpositions of physical Q-shape drive collapse.  How does this yield a causal role for consciousness?

On a materialist view which identifies physical Q-shape (a physical property) with phenomenal Q-shape (a  property of consciousness), the causal role is straightforward.  Superpositions of consciousness involve superpositions of phenomenal Q-shapes, which trigger collapse onto more definite phenomenal Q-shapes, which are themselves more definite physical Q-shapes, leading to more definite physical consequences.

On a dualist view, physical Q-shape may be ontologically distinct from phenomenal Q-shape, so a causal role for the former is not yet a causal role for consciousness.  The simplest way to derive a causal role for phenomenal Q-shape is to assume (i) that consciousness has a quantum structure whereby subjects are in superpositions of phenomenal Q-shapes iff they are in corresponding superpositions of physical Q-shapes, and (ii) a fundamental principle saying that phenomenal Q-shape is super-resistant and obeys the collapse dynamics we have developed.  When subjects are in superpositions of phenomenal Q-shapes, these Q-shapes collapse according to the dynamics.  Phenomenal Q-shapes are perfectly correlated with physical Q-shapes, so collapse of phenomenal Q-shapes leads to collapse of physical Q-shapes, and the standard ensuing physical effects of collapse.

Someone might object that we do not give a genuine causal role to nonphysical consciousness at all.  Instead, all the causal work is done by the physical correlates of consciousness.

One version of this objection notes that on a dualist consciousness-collapse interpretation, there will be PCC states (e.g. physical Q-shapes, on the IIT framework) that correlate perfectly with consciousness.  One can then develop a physicalist collapse interpretation on which the primary locus of superposition-resistance is the PCC states.  Collapse of the PCC states does all the causal work, and collapse of consciousness is causally irrelevant.  There will at least be a possible world (we might think of it as a quantum zombie world) where collapse works this way.  In that world, the physical wave function will evolve just as in our world.  So even in our world, consciousness may seem redundant.

In response: on the dualist interpretation spelled out above, it is consciousness that directly causes the wave function to collapse. There is a fundamental principle saying that consciousness resists superposition. (In the IIT framework, phenomenal Q-shapes resist superposition.) This leads to probabilistic collapse toward determinate states of consciousness.  This collapse of consciousness brings about physical collapse to a more determinate PCC state, because of a psychophysical law ensuring that states of consciousness and their physical correlates (in the IIT framework, phenomenal Q-shapes and physical Q-shapes) are always in alignment. So consciousness is causally responsible for collapse in our world.   There may be other models where physical correlates cause collapse directly, but that is not how things work on the dualist interpretation we have specified.  

The quantum zombie scenario does suggest that there is a sort of structural/mathematical explanation that might be given for our actions without mentioning consciousness.  Still (as is familiar from discussions of panpsychism and Russellian monism), this structural explanation would not provide a complete explanation of our actions, precisely because it leaves out the role of consciousness in grounding that structure.  Like many structural explanations, it leaves out the actual causes.  In the actual world consciousness is causing the relevant behavior, and consciousness may explain why it is that we behave determinately at all.  

A related objection asks: in the actual world, how do we know that it is consciousness that triggers collapse, and not its physical correlates?  As we discussed in the last section, if there is a perfect correlation between the two, these hypotheses cannot be distinguished experimentally.  Still, insofar as we already have reason to believe that consciousness is a fundamental property, then the hypothesis that consciousness triggers collapse has at least two advantages.  First, this way the fundamental law of collapse involves a fundamental property.  Second, this way we have a causal role for consciousness, cohering with a strong pretheoretical desideratum.  These virtues give reasons to favor the view over the alternative.

One might also object that even if our models give consciousness a causal role, they do not give consciousness the kind of causal role that we pretheoretically would expect it to have.  One worry is that collapsing consciousness may affect the objects we perceive, but we want consciousness to affect action, producing intelligent behavior and verbal reports such as ‘I am conscious'.

One worry is that the most obvious effects of collapse point the wrong way: collapse of consciousness will collapse perceived objects such as measurement instruments, but what we want is for consciousness to affect action.  In response, we can note that a collapse of consciousness will collapse an associated PCC state in the brain, and this brain state will be entangled with action states or will at least cause a corresponding action state, so a collapse of consciousness will help bring about a determinate action.  For example, if consciousness probabilistically collapses into an experience of red rather than an experience of blue, this collapse will bring about a PCC state associated with experience of red, which will tend to lead to an utterance of 'I am experiencing red' rather than 'I am experiencing blue'.

Furthermore, consciousness also involves the experience of agency and action: say, the experience of choosing to lift one's left hand rather than one's right hand.  Superpositions of these states will collapse into definite states, which will lead to actions such as raising one’s left hand.  

This picture naturally raises issues about free will.  On this view, the experience of choice plays a nondeterministic causal role in bringing about action.  On some popular conceptions of ``free will", on which what matters for free will is nondeterminism and a role for consciousness, this picture may vindicate free will in the relevant sense.  Others may object that the choices are themselves selected probabilistically, and that random choices are no better than deterministic choices when it comes to free will.  We think the issues are far from straightforward, so we will set aside issues about free will here, but we note that a causal role for consciousness can be expected to have some bearing on those issues.

Another objection is that if consciousness always collapses via the Born role, then any effect of consciousness on action will at best be a sort of dice-rolling role.  It will probabilistically select between different available outcomes, but it will not yield a qualitatively special outcome.  Under a hypothesis where PCC states collapse the wave function, purely physical quantum zombies would have behaved the same way.  So consciousness will not make outcomes on which humans behave intelligently or on which they say 'I am conscious' any more likely than they would have been if some other property had collapsed the wave function.  One might even simulate the dynamics in a classical computer (with a pseudorandom number generator), with no role for consciousness, and the same patterns of behavior would ensue.

Most of what this objector says is correct.  The quantum zombie scenario suggests that there is a sort of structural/mathematical explanation that might be given for our actions without mentioning consciousness.  Still, this structural explanation would not provide a complete explanation of our actions, precisely because it leaves out the role of consciousness in grounding that structure.  (Like many structural explanations, it leaves out the actual causes.)  In the actual world consciousness is causing the relevant behavior, and consciousness may explain why it is that we behave determinately at all.  One might have liked a stronger, more transformative causal role for consciousness that could not even in principle have been duplicated without consciousness, but it is not clear why such a role is essential.

If one does want a stronger role for consciousness, the most obvious move is to suggest that the role for consciousness in collapse is not entirely constrained by the Born probabilities.  Perhaps perceptual consciousness obeys those constraints (thereby explaining our observations in quantum experiments), but agentive experience does not.  For example, collapses due to agentive experience might be biased in such a way that more ``intelligent" choices that lead to more intelligent behavior tend to be favored than they would be according to the Born rule.  This picture sacrifices the great simplicity of the original quantum dynamics, and it could perhaps be disconfirmed through the right sort of experiments and simulations, but it is arguable that our current evidence leaves room open for it.  We do not find this picture especially attractive, but it is at least worth putting it onto the table.\\

\section{Philosophical objections}

We have already considered many objections to our account.  Some are technical issues specific to the use of IIT: for example, whether IIT applies to real physical states, whether Q-shape operators can be defined, and whether a Q-shape/collapse theory has already been falsified by existing experimental results.  These are serious issues that may require modifying IIT or moving to a different theory of the physical correlates of consciousness.  Some are versions of objections that arise for many objective collapse theories: for example, consistency with relativity and the tails problem.  These are also serious issues that we have set aside for now with the preliminary aim of getting consciousness-collapse models closer to the level of seriousness of existing objective collapse theories.  A final technical issue is whether the parameters of a consciousness-collapse theory can be set to avoid the Zeno effect.   

In this final section we consider a number of philosophical objections.  We have already considered objections concerning the causal role of consciousness.  The largest objection remaining concerns superposed states of consciousness.\\

\noindent
\textbf{Objection 1: What is a superposed state of consciousness?}

As we saw earlier, Wigner said that it is “absurd” to suppose that a subject could be in a state of “suspended animation”, that is, in a superposition of multiple states of consciousness.  However, the approximate super-resistance model we have developed requires that subjects can be in such superposed states.   Large superpositions of consciousness (those between significantly different states with significant amplitude for significant periods) will be rare, at least on a fast-collapse model, but they will be possible.  Small superpositions of consciousness (those that are like large superposition except that they are brief, or low-amplitude, or between closely related states) may be ubiquitous.  In fact, on these models it may be that most or all conscious subjects are in small superpositions of consciousness most or all of the time.  This raises the questions: are superpositions of consciousness possible, and if so how can we understand them?

There are a few different ways of trying to understand superposed states of consciousness.  First, one could try to understand them as familiar states: for example, a superposition of seeing an object at positions A and B might be a state of double vision.  However, double vision is an ordinary state of consciousness that can enter superpositions.  It leads to reports such as “I see an object at A and at B”.  The superposed state does not.  It leads to reports such as “I see an object at A”  (if the introspection and report process triggers collapse), or at worst a superposition of “I see an object at A” and “I see an object at B” (if no collapse is triggered). This brings out that the sort of superpositions we need are not introspectible or reportable and will be quite different from familiar states such as double vision.\footnote{\textcite{Shimony1963} reads \textcite{LondonBauer1939} as allowing superpositions of consciousness and critiques the idea in part by arguing that phenomena such as blurred vision and indecision do not really involve superpositions.} 

A more radical alternative says that superposed states of consciousness involve multiple subjects having distinct total states of conscious experience. We will set aside this option as extravagant (do subjects pop into and out of existence in superposition and collapse?), though it is perhaps worth some attention.

A third option is to say that a superposition of states of consciousness is a state that the subject is in, but it is not itself a total state of consciousness. That is, when a subject is in a superposition of conscious states A and B, there is no subjective experience of being in this superposition. There is something it is like to be in A, and something it is like to be in B, but nothing it is like to be in A and B simultaneously.  The subject has the experience of being in A and the experience of being in B, without having any conjoint experience of being in the superposition.  This violates the Unity Thesis articulated by \textcite{BayneChalmers2003} holding that whenever a subject is in multiple conscious states, they are also in a single conscious state that subsumes and unifies them. Some theorists hold that the Unity Thesis is false, at least for split-brain patients and other fragmented subjects: these subjects do not have a single determinate total conscious state, but instead have multiple conscious states as fragments.\footnote{On split-brain cases, see for example \textcite{Nagel1971} who argues for indeterminacy here. \textcite{BayneChalmers2003} argue that in these cases there is a single subject with a single determinate state of consciousness, while \textcite{Schechter2017} argues that there are multiple subjects each with a determinate state of consciousness.}  It is far from obvious what is really going on in these cases, and any analogy with superposed states seems fairly distant. Still, these cases at least bring out that the Unity Thesis and the corresponding assumption that every subject is in a single determinate total state of consciousness is not non-negotiable.

A fourth option is to say that a superposition of total states of consciousness is itself a total state of consciousness – albeit one quite unlike the ordinary total states of consciousness that we are introspectively familiar with.  On this view, when a subject is in a superposition of conscious states A and B, there is something it is like to be in this superposition.  It presumably involves some combination of the experience of being in A and the experience of being in B, combined by some novel phenomenal mode of combination.  This mode of combination is not something we could introspect or report for the reasons discussed above, so it would have to be something that we have no introspective familiarity with.  The phenomenological role of amplitudes is also not clear. Perhaps amplitudes give the ordinary states of consciousness relative weights in the combined states.  As a result, it is far from clear what the phenomenology of a superposed state would be like.  Still, it is far from obvious that a mode of combination like this is impossible.

We think that the fourth option is perhaps the most worthy of consideration, followed by the third. On the fourth option, we can no longer say that total states of consciousness correspond one-to-one with PCC eigenstates. Instead, ordinary non-superposed total states of consciousness will correspond to PCC eigenstates, and superposed total states of consciousness will correspond to superpositions of these eigenstates.

There is precedent to the thought that there are states of consciousness that we cannot introspect or report. Theorists (e.g. Block) who believe in an “overflow” of consciousness outside attention often postulate such aspects: if introspecting and reporting a state always involve attending to it, unattended states cannot be introspected or reported.  One can perhaps make unnoticed superpositions more palatable by noting that on a fast-collapse model they will usually be small superpositions, involving very similar states of consciousness, very low amplitudes, and/or very brief periods of time.  As a result, the superpositions may largely fall below the grain of our ordinary introspective access.  

Still, the fact that our super-resistance model has to postulate superposed states of consciousness is a significant cost of the view.  Is it possible to develop a super-resistance consciousness-collapse model that avoids superpositions of consciousness while also avoiding the Zeno problem?  Such a model would need to give up on the tight connection between definite conscious states and PCC eigenstates, in order that never-superposed conscious states do not lead to never-superposed PCC states and so to the Zeno effect.   At the same time, it would need to retain enough of a connection between consciousness and physical states that the definiteness of consciousness leads to collapse in its physical basis.  It is not easy to meet both demands at once. One path invokes a looser connection between consciousness and PCC eigenstates, whereby superposed PCC states can coexist with definite states of consciousness at least briefly. For example, one might hold that superposed PCC states determine a definite state of consciousness probabilistically according to the Born rule, and that this definite state of consciousness leads to collapse onto a corresponding PCC state but only after a time delay.  Perhaps this view and others in the neighborhood are at least worth developing. 

In any case: in ordinary quantum mechanics, many theorists say that they cannot really imagine what it is for a physical state to be in a superposition.  At the same time, they adopt the idea and run with it, and the idea seems to be theoretically fruitful.  Our suggestion is that we do something like this for superpositions of states of consciousness, at least for now.  We should simply adopt the idea and see whether it is fruitful. If it is, we can later return to the question of just what superposed states of consciousness involve.\\

\noindent
\textbf{Objection 2: How do quantum effects make a difference to macroscopic brain processes?}

Quantum theories of brain processes are sometimes criticized on the grounds that it is hard to see how low-level quantum processes can affect high-level processing in neurons.  A more specific version of this objection is that on some accounts (e.g. Hameroff and Penrose), quantum coherence at the neural level is required for distinctively quantum effects in neural processing, but the high temperatures in the brain are likely to lead to decoherence below the neural level.  These objections do not apply to our approach, which does not involve any special effects of low-level quantum processes on neural processes and is entirely consistent with decoherence at relatively low levels.  In fact, in our central illustrations, we have treated brain states as superpositions of numerous decoherent eigenstates, which themselves may involve relatively classical processing in neurons. The only high-level quantum process that plays an essential role in our framework is the collapse process, which selects one or more of these eigenstates as outlined above. Our picture is consistent with further macroscopic quantum effects, but they are not required.\\ 

\noindent
\textbf{Objection 3: What about macroscopic superpositions?}

One might worry that on a consciousness-collapse view ordinary macroscopic objects such as measurement devices will exist in states of superposition until they are observed.  Our view does not necessarily lead to this consequence. For a start, if a correct theory of consciousness associates these devices with some amount of consciousness (as may be the case for IIT), then the devices will collapse wave functions much as humans do.  Even if these devices are not conscious, it is likely that typical measuring devices will be entangled with humans and other conscious systems, so that they will typically be in a collapsed state too.  Still, in special cases where such a device is entirely isolated from conscious systems and records a quantum interaction, it will enter a macroscopic superposition.  Of course we will never observe such a superposition, as our observation will collapse the state of the system.  But we might in principle get empirical evidence of this superposition if we can eventually measure associated interference effects.  Perhaps the existence of macroscopic superpositions is counterintuitive, but many cosmological theories already allow macroscopic objects to be in superposition in the early universe where there are no observers.  It is unclear why allowing this in the current universe is any worse. \\

\noindent
\textbf{Objection 4:  What about the first appearance of consciousness in the universe?}

As we saw earlier, if consciousness is absolutely super-resistant, the quantum Zeno effect entails that it can never emerge for the first time in the development of the universe. On an approximate super-resistance model, there is less of a problem.  For eons, the universe can persist in a wholly unconscious superposed state without any collapses.  At some point, a physical correlate of consciousness may emerge in some branch of the wave function, yielding a superposition of consciousness and unconsciousness (or their physical correlates) with low amplitude for consciousness.  With high probability the universe will collapse back toward an unconscious state.  As this happens repeatedly in many branches of the wave function, there will eventually be a low probability collapse toward a state of consciousness, and consciousness will be in a position to take hold.

\section{Conclusion}

The results of our analysis are mixed.  We have developed a consciousness-collapse model with a reasonably clear and precise dynamics.  But it must be admitted that the model we have developed is not as simple and powerful as the original (simple if imprecise) measurement-collapse framework.

Our initial superselection collapse model was simple, but it leads to the Zeno problem.  Avoiding the Zeno problem has led to a number of complications. First, we have had to countenance superpositions in states of consciousness, and it is not at all clear that this is possible.  Second, we have had to introduce Pearle-style collapse dynamics along with parameters for the rate of collapse, and these parameters have to be constrained carefully in order to yield empirically acceptable results.  We have also had to invoke a complex theory of consciousness -- though this is less of a cost, since a theory of consciousness is needed even in the absence of the quantum measurement problem.

Is this consciousness-collapse model the best that we can do?  We have seen that to avoid countenancing superposed states of consciousness while also avoiding the Zeno problem, a consciousness-collapse model will need to break the strong link between definite states of consciousness and eigenstates of a PCC observable.  Perhaps there are alternative models on which the physical correlates of consciousness involve a more complex wave-function property, or on which consciousness can vary independently of any physical properties.  There also remain the possibility of variable-locus models, though these may also need to break the strong link between consciousness and its physical correlates to avoid the Zeno problem.  In any case, models along these lines are certainly worth exploring.

Overall: the model we have developed is perhaps not as simple or powerful as some of the leading interpretations of quantum mechanics. If it is the best we can do, then the upshot may be that consciousness-collapse models are subject to principled limitations.  Nevertheless, it at least serves as an existence proof for a relatively precise consciousness-collapse model.  The model is open to empirical test, and it is not out of the question that a more powerful model along these lines could be developed.  In the meantime, the research program of consciousness-collapse models deserves attention.

\appendix

\section{Appendix: Calculating Q-shape for a dyad system in IIT 3.0 and quantum IIT}

In this appendix, we illustrate some mathematical details of standard IIT (IIT3.0) and quantum IIT (QIIT), by showing how $\Phi$ and Q-shape are determined in simple dyad systems with two elements. The IIT formalisms are complex, but dyads avoid some  complications. We will also define a distance measure between Q-shapes which is important for the collapse dynamics.   

We begin with IIT3.0.\footnote{Thanks to Nao Tsuchiya and Leo Barbosa. Our calculations follow the supporting information in \textcite{Mayneretal2018} especially S1: Calculating $\Phi$. See also \textcite{Oizumietal2014} and \textcite{Tononietal2016}. For the earlier, simpler IIT formalism for calculating $\Phi(AB)$, see \textcite[fig. 5]{Tononi2004}, \textcite{Tsuchiya2017}, and \textcite{McQueen2019b}. The reader can experiment with calculating $\Phi$ for various systems including the dyad AB at http://integratedinformationtheory.org/calculate.html. Details of the underlying software can be found in \textcite{Mayneretal2018}.} We assume a dyad system with two elements A and B, each of which can be in one of two states: [1] or [0]. The composite system AB can be in one of four possible states: [11], [00], [10], or [01].  The transition rules are a simple swap: the state of A at one time is determined by copying the state of B at the previous time and vice versa.  We can stipulate that in the system under consideration, the current state of AB is [10].  The next state is thereby determined to be [01]. Subsystems of AB are the nonempty sets of elements of the system: \{A\}, \{B\}, and \{A, B\}, which we will abbreviate as A, B, and AB when there is no chance of confusion. Mechanisms are subsystems with nonzero weight.

The Q-shape of a system consists of a location $L(m)$ for each mechanism $m$ in the system, weighted by the measure $\phi(m)$. $L(m)$ is a point in a $2^{n+1}$-dimensional space with two dimensions for each of the $2^n$ possible states of the system, where $n$ is the number of elements. $L(m)$ is determined by conjoining two probability distributions over the states $S$ of the system: $p_m(S)$ and $p'_m(S)$, where the former is defined in terms of the effects of $m$ and the latter is defined in terms of the causes of $m$. Each distribution is associated with a $\phi$ value. The weight $\phi(m)$ is the minimum of these two values. The Q-shape of AB lives in an 8-dimensional space, as AB has four possible states. As we will see, of the three subsystems of AB, only A and B yield mechanisms with nonzero weight. Hence, The Q-shape of AB consists in two weighted points located in an 8-dimensional space.


IIT3.0 distinguishes two notions of integrated information: $\phi$ (small phi), which applies to individual mechanisms, and $\Phi$ (big phi), which applies to the total system.  To know $\Phi(AB)$ we must first calculate AB's Q-shape. To know AB's Q-shape we must first calculate $\phi$ for AB’s mechanisms. To begin with, we illustrate how the  probability distribution $p_m(S)$ and $\phi(m)$ are calculated and then used to define Q-shape and $\Phi$.

The distribution $p_m(S)$ is a distribution over future states $S$ of the system, reflecting their probability of occurrence given that the elements of $m$ are fixed to their current state (while any other elements are allowed to vary).  For the candidate mechanism AB, both elements will be fixed to their current value [10].  $p_{AB}(S)$ is the probability that the following state will be S, given the current state [10]. The following state is guaranteed to be [01], so $p_{AB}$ assigns probability 1 to [01] and probability 0 to the other three states.

Recall that $L(m)$ is determined by conjoining two probability distributions over the states $S$ of the system: $p_m(S)$ and $p'_m(S)$. If we consider just $p_m(S)$, then the location $L(AB)$ can be seen as a point in 4-dimensional space corresponding to the distribution $p_{AB}(S)$. Let us say the four dimensions are ordered as [00], [01], [10], [11].  Then L(AB)=[0,1,0,0], which assigns 1 to [01] and 0 to the other states. While $p_m(S)$ is a distribution over possible future states, $p'_m(S)$ is a distribution over possible preceding states  (i.e. the probabilities that the preceding state was $S$, given the current state). For our system AB the two distributions are the same, so the 8-dimensional location will be a repeated version of the 4-dimensional location: that is, L(AB)=[0,1,0,0,0,1,0,0].

For candidate mechanism A, $p_A(S)$ is the probability of the following state being $S$ given that element A is fixed to its current value [1], while the other element B can vary with probability 0.5 for each value [0] or [1]. Under these conditions, the following state may be either [01] or [11], and $p_A$ will assign these two states probability 0.5 each. Likewise, $p_B$ will assign probability 0.5 each to states [00] and [01], the two states that can follow a state where B is fixed to 0. As with AB, $p_m(S)$ = $p'_m(S)$ for A and for B. As a result, $L(A)=[0,0.5,0,0.5,0,0.5,0,0.5]$ and $L(B)=[0.5,0.5,0,0,0.5,0.5,0,0]$.

The integrated information [small phi] $\phi(AB)$ is determined by considering the difference between the probabilistic effects of the subsystem AB with the effects of a partitioned subsystem A-B where we consider only the effects of A and B taken separately on each other.  We can define a probability distribution $p_{A-B}$ as the tensor product of two distributions: a distribution $p_{A|B}$ over states of A given that B is fixed to its current value 0 (so A=[1] has probability 1) and a distribution $p_{B|A}$ over states of B given that A is fixed to its current value 1 (so B=[0] has probability 0).  The product distribution $p_{A-B}$ assigns 1 to [10] and 0 to every other state.

We can then define $\phi(AB) = EMD(p_{AB}, p_{A-B})$. For two probability distributions $p_1$ and $p_2$ over the same state-space, EMD($p_1,p_2$) is the Earth mover’s distance between $p_1$ and $p_2$.  This can be defined as the minimal amount of work required to turn $p_1$ into $p_2$ by moving the ``Earth" of probability from some points in the $2^n$-dimensional space to other points, where work is measured by the amount of probability moved multiplied by the Hamming distance between the points. In the case just described, $p_{AB}$ and $p_{A-B}$ are exactly the same distribution, so the Earth mover’s distance between them is 0. So $\phi(AB) = 0$. 

The quantity $\phi(A)$ can be defined as a related Earth-mover’s distance over states of B, comparing the distribution over those states with A fixed to its current value of [1] (resulting in probability 1 to B=[0]) to a distribution that ignores the value of A (resulting in probability 0.5 each to B=[0] or B=[1]). In this case, $\phi(A) = 0.5$. Likewise, $\phi(B)=0.5$.

As a result, we can fully specify the Q-shape $Q_{AB}$ of the system AB. It consists of location $L(AB)=[0,1,0,0,0,1,0,0]$ with associated weight $\phi(AB) = 0$, location $L(A)= [0,0.5,0,0.5,0,0.5,0,0.5]$ with associated weight $\phi(A)=0.5$, and location $L(B)=[0.5,0.5,0,0,0.5,0.5,0,0]$ with associated weight $\phi(B)=0.5$.

In the above calculations we took a shortcut that should now be made explicit. For each candidate mechanism, we chose to consider the probability distribution assigned to the future (or past) states \textit{of a particular subsystem}. For candidate mechanism AB we chose subsystem AB. For candidate mechanism A we chose subsystem B. And for B we chose A. These choices are not arbitrary, but are the result of an optimization procedure. For each candidate mechanism, we in fact consider all possible subsystems and choose the subsystem that maximizes $\phi$. For example, when considering candidate mechanism A, it turns out that A has more integrated information about B than about AB. After all, there are three possible ways of disconnecting A from AB: disconnect A to A, A to B, A to AB. Nothing happens by disconnecting A to A (there was no connection there to begin with!). But then that is the minimal information partition, implying that A has zero $\phi$ about AB. On the other hand, there is only one way to disconnect A from B, and that disconnection does make a difference, giving nonzero $\phi$. For details see \textcite{Barbosaetal2021}.

We can define the distance between two Q-shapes $Q_1$ and $Q_2$ (defined over the same states $S$, with associated probability distributions $p_{m,1}$ and $p_{m,2}$ and weights $\phi_1(m)$ and $\phi_2(m)$) as an extended Earth mover’s distance $EMD^*(Q_1, Q_2)$:

\begin{equation}\label{EMD*}
\begin{split}
EMD^*(Q_1, Q_2) = \\
\sum_i (|\phi_1(m_i)-\phi_2(m_i)|\times  (EMD(p_{m_i,1} , p_{m_i,2})) + EMD(p'_{m_i,1} , p'_{m_i,2})))
\end{split}
\end{equation}


This distance is the minimal amount of work required to transform the $\phi_1$ distribution over mechanisms $m$ into $\phi_2$ by repeatedly moving the ``Earth" of $\phi$ from one mechanism $m_1$ in $Q_1$ to another mechanism $m_2$ in $Q_2$. (A complication is that in some cases (where $Q_1$ has more total $\phi$ than $Q_2$), we need to send the excess to an unconstrained distribution $p_{uc}$ associated with $Q_2$.)


We can then define $\Phi(AB)$ as the minimal value of EMD*($Q_{AB}, Q_{AB^*})$, across all partitions $AB^*$ of $AB$.  A partition of a system requires cutting one or more causal connections between its units.  For system AB, a partition cuts the connection from A to B or from B to A or both.  In this case, either cut reduces $\phi$ to zero for both mechanisms A and B, and their probability distributions are flattened. The reason why cutting just one of these two connections destroys both mechanisms is tied to the fact that $\phi(m)$ is defined as the minimum of two $\phi$ values, the one that pertains to the future state and the one that pertains to the past state. Each cut will send one of these $\phi$ values to zero.

Recall that $Q_{AB}$ assigns $\phi(A) = \phi(B)=0.5$, where these serve as weights for $L(A)=[0,0.5,0,0.5,0,0.5,0,0.5]$ and $L(B)=[0.5,0.5,0,0,0.5,0.5,0,0]$. $Q_{AB^*}$ instead assigns zero weights to both $L(A^*)$ and $L(B^*)$, where $L(A^*) = L(B^*) = [0.25,0.25,0.25,0.25,0.25,0.25,0.25,0.25]$. We thus have:
$EMD^*(Q_{AB}, Q_{AB^*}) = |\phi(A)-\phi(A^*)|\times  (EMD(p_A, p_{A^*}) + EMD(p'_A, p'_{A^*})) +  |\phi(B)-\phi(B^*)|\times  (EMD(p_B, p_{B^*}) + EMD(p'_B, p'_{B^*}))$ = $(0.5\times  (0.5 + 0.5)) + (0.5\times  (0.5 + 0.5)) = 1$.


The crucial quantity $\Phi^{max} (AB)$ is defined as $\Phi(AB)$ if AB is a maximum of $\Phi$, and 0 otherwise.  Here AB is a maximum of $\Phi$  if $\Phi(AB) > \Phi(S)$ for all systems $S$ such that $S$ has elements in common with AB.  In our case, we can stipulate that AB is isolated from its environment so that no other system containing A or B has higher $\Phi$.  In this case, AB is a maximum of $\Phi$, so $\Phi^{max}(AB) = 1$. According to IIT, $\Phi^{max}$ is a measure of consciousness, so system AB has one unit of consciousness.

In section 4 we noted that if AB is in a different state (either 01, 00, or 11), than the calculation for $\Phi$ is the same, but the Q-shape is different. This can now be seen by the fact that changing the initial state changes the locations but not their weights. Thus, if the initial state is instead 00, then we still have two mechanisms A and B, each with weight 0.5, but their locations become $L(A)=[0.5,0.5,0,0,0.5,0.5,0,0]$ and $L(B)=[0.5,0,0.5,0,0.5,0,0.5,0]$. This is not enough to change $\Phi$, but it is enough to change the Q-shape. We can thus define  Schroedinger's dyad as AB in a superposition of 10 and 00. A collapse model base only on $\Phi$ would fail to collapse this superposition, despite it being a superposition of conscious states. 

We now move to quantum IIT (QIIT).\footnote{Thanks to Johannes Kleiner. Our calculations are intended to follow \textcite{zanardi2018quantum} and \textcite{KleinerTull2020}.} To simplify the calculations of the dyad, it is easier to start A and B in the same initial state ($\ket{00}$ or $\ket{11}$) so that they remain stationary. We add the further stipulation that A (B) maintains its own state over time. We may now consider A and B to be AND gates that each take two inputs as depicted. 

\begin{center}
\includegraphics[scale=.7]{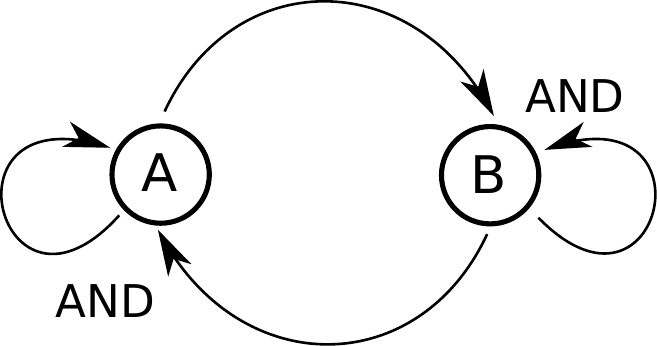}\\[.5em]
\end{center}

State $|00\rangle$ has zero $\Phi$ and Q-shape, that is, $\Phi(|00\rangle) = Q(|00\rangle) = 0$. For if we partition the system by replacing one of the directed edges with random input, the inputs are still only either 00 or 01, whereas the AND gates require an input of 11 to change state. Partitioning does not make a difference. 

Partitioning makes a difference if the system is instead in state $|11\rangle$: If any of the edges are removed and replaced by random input, at least in half the cases it will feed a $0$ to its target, so that in light of the AND gate the state of the target will change from $1$ to $0$. This implies that the system in that state has non-zero $\Phi$ value, and its Q-shape isn't null. 

We can therefore introduce collapse operators for the Q-shapes of these two states, and then use them to define a small consciousness superposition. 

Our new dyad still has three subsystems (AB, A, and B). For each we consider the integrated information $\phi$ of both future and past states. So for the collapse operators $Q^k_{ij}$, the $k$ index runs from 1 to 6. Since the Hilbert space of the system in this case is $4$ dimensional, the indices $i$ and $j$ run from $1$ to $4$ each.

The $c_{ij}^k(\psi)$ in (\ref{Kleiner}) are the coefficients of the operator $\rho$ which is the $k$th component of the Q-shape of $\psi$. Because $Q(00) = 0$, it follows that $c_{ij}^k(\psi_0)= 0$. 
Since $\ket{00}$ and $\ket{11}$ are wave functions with classical Q-shapes, they are contained in $\mathds{E}^{-1}(\mathcal C)$ and are summed over in (\ref{Kleiner}). It follows that

\begin{equation}
    \hat Q^k_{ij} \ket{00}  =  \sum_{\psi \in \mathds{E}^{-1}(\mathcal C)} \varphi^k(\psi) c_{ij}^k(\psi) \ket{\psi} \bra\psi\ket{00} = \varphi^k(00) c_{ij}^k(00)\ket{00} = 0\ket{00} \end{equation}

We have assumed the wave functions with classical Q-shapes are orthogonal. Thus $\ket{00}$ is an eigenvector of every operator $\hat Q^k_{ij}$ with eigenvalue $0$. We also have 

\begin{equation}
    \hat Q^k_{ij} \ket{11}  =  \sum_{\psi \in \mathds{E}^{-1}(\mathcal C)} \varphi^k(\psi) c_{ij}^k(\psi) \ket{\psi} \bra\psi\ket{11} = \varphi^k(11) c_{ij}^k(11)\ket{11} \end{equation}

so that $\ket{11}$ is an eigenvector of $Q^k_{ij}$ with eigenvalue $c_{ij}^k(\ket{11})$. 

Letting $\ket{11}$ be the alive (conscious) state and $\ket{00}$ be the dead (unconscious) state, we can provide (in addition to the section 4 example) another example of Schroedinger's dyad:

\begin{equation}
    \Ket{\Psi}_{AB} = \alpha\Ket{00} + \beta\Ket{11}
\end{equation}

Our dynamics (in section 6) predicts that this state is not completely stable, but continuously collapses towards one of the two Q-shape eigenstates, in accord with the Born rule.


\printbibliography


\end{document}